# The influence of funding on the Open Access citation advantage

Pablo Dorta-González [1,*] and María Isabel Dorta-González [2]

[1] Universidad de Las Palmas de Gran Canaria, TiDES Research Institute, Campus de Tafira, 35017 Las Palmas de Gran Canaria, Spain. E-mail: pablo.dorta@ulpgc.es ORCID: http://orcid.org/0000-0003-0494-2903

[2] Universidad de La Laguna, Departamento de Ingeniería Informática y de Sistemas, Avenida Astrofísico Francisco Sánchez s/n, 38271 La Laguna, Spain. E-mail: isadorta@ull.es

* Corresponding author

## Abstract

Some of the citation advantage in open access is likely due to more access allows more people to read and hence cite articles they otherwise would not. However, causation is difficult to establish and there are many possible bias. Several factors can affect the observed differences in citation rates. Funder mandates can be one of them. Funders are likely to have OA requirement, and well-funded studies are more likely to receive more citations than poorly funded studies. In this paper this hypothesis is tested. Thus, we studied the effect of funding on the publication modality and the citations received in more than 128 thousand research articles, of which 31% were funded. These research articles come from 40 randomly selected subject categories in the year 2016, and the citations received from the period 2016-2020 in the Scopus database.

We found open articles published in hybrid journals were considerably more cited than those in open access journals. Thus, articles under the hybrid gold modality are cite on average twice as those in the gold modality. This is the case regardless of funding, so this evidence is strong. Moreover, within the same publication modality, we found that funded articles generally obtain 50% more citations than unfunded ones. The most cited modality is the hybrid gold and the least cited is the gold, well below even the paywalled. Furthermore, the use of open access repositories considerably increases the citations received, especially for those articles without funding. Thus, the articles in open access repositories (green) are 50% more cited than the paywalled ones. This evidence is remarkable and does not depend on funding. Excluding the gold modality, there is a citation advantage in more than 75% of the cases and it is considerably greater among unfunded articles. This result is strong both across fields and over time.

**Keywords:** open access; funded research bias; gold OA; green OA; hybrid OA; scholarly communication; science policy.

## Introduction

Researchers are more likely to read and cite papers to which they have access than those that they cannot obtain. Thus, since the emergence of the world wide web, scientists and scholarly publishers have used different forms of Open Access (OA), a disruptive model for the dissemination of research publications (Björk, 2004). In the last years, more and more scientists are making their research results openly accessible to increase its visibility, usage, and citation impact (Dorta-González et al., 2017; 2020).

The common characteristic of all different forms of OA is that the primary source of communication of research results, the peer reviewed article, is available to anybody with Internet access free of charge and access barriers (Prosser, 2003).

Thus, there are four main OA modalities. *Gold OA* refers to scholarly articles in fully accessible OA journals. *Green OA* refers to publishing in a subscription or pay-per-view journal (paywalled), in addition to self-archiving the pre- or post-print paper in a repository (Harnad et al., 2004). *Hybrid Gold* is an intermediate form of OA, where authors pay scholarly publishers to make articles freely accessible within journals, in which reading the content otherwise requires a subscription or pay-per-view (Björk, 2017). And *Bonze OA* (delayed OA) refers to scholarly articles in subscription journals made available openly on the web directly through the publisher at the expiry of a set embargo period (Laakso & Björk, 2013).

Hybrid journals are a risk free transition path towards full OA (gold), in contrast to starting new full OA journals or converting journals, since the subscription revenue remains (Prosser, 2003). However, the price level in the hybrid gold is typically around 3,000 USD, which many authors and their institutions perceive as high (Tenopir et al., 2017).

Since Lawrence proposed in 2001 the OA citation advantage, this postulate has been discuss in depth without an agreement being reach (Davis et al., 2008). Furthermore, some authors are critical about the causal link between OA and higher citations, stating that the benefits of OA are uncertain and vary among different fields (Davis & Walters, 2011).

Some of the citation advantage in open access is likely due to more access allows more people to read and hence cite articles they otherwise would not. However, causation is difficult to establish and there are many possible bias. Several factors can affect the observed differences in citation rates. Funder mandates can be one of them. Funders are likely to have OA requirement, and well-funded studies are more likely to receive more citations than poorly funded studies (Aagaard et al., 2020).

In this paper, this hypothesis is tested. Thus, based on citation data from the Scopus database, we provide longitudinal estimations of cites per article in all publication modalities: gold, hybrid gold, bronze, green, and paywalled. Moreover, we will answer the following questions:

1. Are OA research articles more highly cited than their paywalled counterparts?

2. Are there differences attributable to financing?

3. Which publication modality brings a greater citation advantage?

4. How does this citation advantage vary according to field and time?

## Theoretical framework on open access citation advantage

Many researchers, starting with Lawrence (2001), have found that OA articles tend to have more citations than pay-for-access articles. This OA citation advantage has been observed in a variety of academic fields including computer science (Lawrence, 2001), philosophy, political science, electrical & electronic engineering, and mathematics (Antelman, 2004), physics (Harnad et al., 2004), biology and chemistry (Eysenbach, 2006), as well as civil engineering (Koler-Povh et al., 2014).

However, this postulate has been discussed in the literature in depth without an agreement being reached (Davis et al., 2008; Dorta-González & Santana-Jiménez, 2018; Norris et al., 2008; Joint, 2009; Gargouri et al., 2010; González-Betancor & Dorta-González, 2019; Wang et al., 2015). Furthermore, some authors are critical about the causal link between OA and higher citations, stating that the benefits of OA are uncertain and vary among different fields (Craig et al., 2007; Davis & Walters, 2011).

Kurtz et al. (2005), and later other authors (Craig et al., 2007; Moed, 2007; Davis et al., 2008), set out three postulates supporting the existence of a correlation between OA and increased citations. (1) OA articles are easier to obtain, and therefore easier to read and cite (*Open Access postulate*). (2) OA articles tend to be available online prior to their publication and therefore begin accumulating citations earlier than pay-for-access articles (*Early View postulate*). And (3) more prominent authors are more likely to provide OA to their articles, and authors are more likely to provide OA to their highest quality articles (*Selection Bias postulate*). Moreover, these authors conclude that early view and selection bias effects are the main factors behind this correlation.

Gaule & Maystre (2011) and Niyazov et al. (2016) found evidence of selection bias in OA, but still estimated a statistically significant citation advantage even after controlling for that bias. Regardless of the validity or generality of their conclusions, these studies establish that any analysis must take into account the effect of time and selection bias.

At journal level, Gumpenberger et al. (2013) showed that the impact factor of gold OA journals was increasing, and that one-third of newly launched OA journals were indexed in the Journal Citation Reports (JCR) after three years. However, Björk and Solomon (2012) argued that the economic model is not related to journal impact. This result has been confirmed by Solomon et al. (2013), concluding that articles are cited at a similar rate regardless of the distribution model.

The OA citation advantage is not universally supported. Many studies have been criticised on methodological grounds (Davis & Walters, 2011), and a research using the randomized-control trial method failed to find evidence of an OA citation advantage (Davis, 2011).

However, recent studies using robust methods have observed an OA citation advantage. McCabe & Snyder (2014) used a complex statistical model to remove author bias and reported a small but meaningful 8% OA citation advantage. Archambault et al. (2014) in a massive sample of over one million articles and using field-normalized citation rates, described a 40% OA citation advantage. Ottaviani (2016) reported a 19% OA citation advantage excluding the author self-selection bias and beyond the first years after publication.

In a recent study, Piwowar et al. (2018) used three samples, each of 100,000 articles, to study OA in three populations: all journal articles assigned a DOI, recent journal articles indexed in Web of Science, and articles viewed by users of the open-source browser extension Unpaywall. They estimated that at least 28% of the scholarly literature is OA, and that this proportion is growing mainly in gold and hybrid journals. Accounting for age and discipline, they observed OA articles receive 18% more citations than average, an effect driven primarily by green and hybrid OA.

## Methodology

The database Scopus has new open access filters since the end of 2020, providing information on the modality of open access per article. With this new classification system, users can now filter their results or use specific open access tags, i.e. gold, hybrid gold, bronze, and green.

The source of OA information in Scopus is Unpaywall, an open-source browser extension that lets users find OA articles from publishers and repositories (hold by Impactstory, a non-profit organization).

In this study, 40 subject categories in the Scopus database were randomly select. This is 12% of the subject categories (40 of 334) and 6.5% of the research articles in the Scopus database in 2016. They resulted 12 subject categories from Health Sciences, 7 from Life Sciences, 10 from Physical Sciences & Engineering, and 11 from Social Sciences & Humanities.

For each subject category, the "research articles" in the year 2016 and the citations received by such research articles in the period 2016-2020, were download from the Scopus database (April 28, 2021). This information is shown in the dataset in Annex A.

In relation to the representativeness of the sample, a total of 1,992,035 research articles were index in the Scopus database in 2016, of which 640,032 specifying a funding source (32.1%). During that same year, the selected 40 subject categories included 128,663 research articles, of which 39,675 were funded (30.8%). The representativeness of the sample by publication modality is shown in Table 1. Thus, the size of the sample over the total population, in number of research articles in 2016, varies according to publication modality between 4.8% and 9.2%.

## Results

The prevalence of the publication modality by funding, both in the sample and in the total database, is shown in Table 2. Thus, most of the research articles in the Scopus database in 2016 were paywalled, two out of three unfunded articles (67%) and half of the funded articles (51%). The use of open access repositories (green), is more widespread within the funded group (41%), compared to the group without funding (24%). The prevalence of the gold modality is quite similar, with only one percent point more in the case of funding. This is because some gold journals are also free of charge for the authors. However, the prevalence of the hybrid gold modality within the group with funding is double that of the group without funding. This is due to the authors must pay the publication costs under the hybrid gold modality.

The prevalence of funding by publication modality is shown in Table 3. Note most of the research articles in the Scopus database in 2016 were unfunded (68%), and only one out of three articles were funded (32%). The prevalence of unfunded articles rises to 73% in the paywalled modality. Although it may be surprising that in the gold modality the prevalence of unfunded articles is almost double that of those with financing, this is because, as already indicated, in the database two out of every three articles do not have financing. However, in the rest of the open access modalities, the proportion between both groups, with and without financing, is quite similar. This is despite the fact that, as already mentioned, the group without financing is much bigger in absolute value. Therefore, in the funded group there is a greater concern about offering open access to publications.

The prevalence of the publication modality in the sample by subject category and funding is shown in Figure 1. There are very important differences among subject categories, both in the prevalence of funding and in the open access modality. Thus, while in the humanities the prevalence of funding is around 5%, in the life sciences it exceeds 50% in some cases. Furthermore, while in some social sciences and humanities the prevalence of open access is below 20%, it reaches over 70% in some scientific disciplines.

*Cites per article by funding and publication modality*

The cites per article in 2016-2020 by funding, publication modality, and branch of knowledge are shown in Table 4. The mean is higher than the median in most cases. However, in life sciences just the opposite happens in some publishing modalities. In general, the highest citation averages are reach in life and health sciences, while the lowest citation averages are obtain in social sciences and humanities.

However, is between funding groups where the biggest differences exist. Thus, within the same modality, financed articles generally obtain 50% more citations than non-financed ones. The most cited modality is the hybrid gold and the least cited is the gold, well below even the paywalled. Thus, articles under the hybrid gold modality are cited on average twice as those in the gold modality, and the green articles 50% more cited than the paywalled ones. Both evidences are remarkable. The first is justified because gold journals are younger than hybrids and in most cases do not have the prestige of the latter. The second is a measure of citation advantage in the open access repositories. Furthermore, these relationships are not dependent on funding.

The box diagram for the distribution of cites per article, according to funding and publication modality, is shown in Figure 2. In all publication modalities, cites per article for those in the funded group are clearly higher than the citations in the unfunded group. These average citations for the funded articles are higher both in mean (indicated with the x symbol) and in quartiles of the distribution (box and whisker). Note that the mean of the distribution is in most cases larger than the median. This is because the distribution is asymmetric with a long tail on the right.

Regardless of funding, open articles published in hybrid journals were considerably more cited than those published in open access journals. Note that 75% of the articles published during 2016 in open access journals (gold) received an average number of citations less than that received by the 25% least cited of open access articles in hybrid journals (hybrid gold). This is the case regardless of funding, so this evidence is strong. However, as mentioned before, gold journals are younger than hybrids and in most cases do not have (at the moment) the prestige of the latter.

With the exception of open access journals (gold), the rest of open access modalities received more citations than paywalled articles. Moreover, the open access modality that receives more citations is the hybrid (hybrid gold). Both results are obtain regardless of funding, so this evidence is also strong.

In the group of unfunded articles, the average citation received by those deposited in open access repositories but published in the paywalled modality (only green), is greater than the average citation of all articles with versions in repositories (green). However, this does not happen within the group of those financed. This means that the use of open access repositories considerably increases the citations received, especially for those publications without funding.

The trend over time of cites per article by funding and modality is shown in Figure 3. Notice the increase in the number of citations over time to a large degree relates to the shape of the citation distribution. Thus, beyond this logical increase in the number of citations over first years after publication, no clear time effect observes in Figure 3.

*Open Access citation advantage*

The OA citation advantage (disadvantage if negative) for an OA modality (gold, hybrid gold, bronze, green, and only green) in a particular year, is defined in relation to the paywalled modality as the difference of citations. If cites per OA article in a particular modality is greater than cites per paywalled article, then the OA citation advantage of that modality is:

(Cites per OA - Cites per paywalled) / Cites per paywalled.

However, if cites per OA article in a particular modality is less than cites per paywalled article, then the OA citation advantage (disadvantage because it is negative) of that modality is:

(Cites per OA - Cites per paywalled) / Cites per OA.

The average OA citation advantage by funding, publication modality, and branch of knowledge is shown in Table 5. Notice the outliers observed in the data distribution can skew the mean. Thus, the median is more robust measure of central tendency than the mean. Half of the articles have OA citation advantage above the median of the distribution and the other half below.

There are important differences between branches of knowledge. For the aggregate of all subject categories and excluding the gold modality, the average OA citation advantage varies in the funded group in the range 41–79%, with a median in 22–69%. In the unfunded group (excluding gold), the OA citation advantage varies in the range 72–124%, with a median in 42–80%.

The highest advantage reaches in hybrid gold, with 79% and 124% for funded and unfunded, respectively. Half of the categories analyzed present hybrid gold citation advantages greater than 69% for funded and 80% for unfunded articles. In green modality, the average OA citation advantage for funded and unfunded articles is 45% and 81%, respectively. Moreover, half of the categories present green citation advantages greater than 35% for funded and 50% for unfunded. In the only green modality, the average OA citation advantage is 47% for funded and 72% for unfunded articles, although half of the categories present advantages greater than 22% and 61%, respectively.

Thus, we can conclude that, excluding the gold modality where there is no OA citation advantage, the citation advantage of the other OA modalities in relation to the paywalled is on average greater than 50% increase in the group of unfunded articles. In half of unfunded articles (median), citation advantages were obtained above 80% in hybrid gold and 50% in green. However, in half of funded articles citation advantages were obtained above 69% and 35% for the hybrid gold and green, respectively.

The distribution of the OA citation advantage in relation to the modality and funding is shown in Figure 4. Note there are differences in OA citation advantage both between funding groups and among OA modalities. The OA citation advantage is clear for all open access modalities, with the exception of the open access journals (gold) as mentioned. The data distribution, represented by the box and whisker, displaces toward the positive part of the vertical axis. Note the range of variation is considerably lower in the funded group. The median of the distribution is the inner line that divides the box into two parts, and the mean is the x symbol. Excluding the gold modality, there is a citation advantage in more than 75% of the cases (the 25th percentile is the bottom line of the box), although something less in the unfunded bronze group.

The OA citation advantage is held in time (see Figure 5). In the gold modality, and regardless of funding, although there is a clear citation disadvantage with respect to the paywalled option, this disadvantage decreases over time. However, in the hybrid gold and bronze modalities, where there is a clear citation advantage in relation to the paywalled, this advantage varies over time without a clear trend, and we can assume that it does not depend on time. The modalities in which the trend is more stable according to the median of the distribution are green and only green.

Finally, the OA citation advantage is consistent across subject categories and held in time (see Figure 6). If we discard the lines further away, which are infrequent, a certain increasing trend can be observed in the gold modality, although most of the lines fall in the negative zone as already noted. There is great variability in the lines of the hybrid gold and bronze options. However, in general it seems that the citation advantage is held in time. In the green and only green groups, with some exceptions, the citation advantage is maintain over time (see also moving average smoothing in Figure 7).

## Conclusions

The access to academic literature is a current debate in the research community. Research funders are increasingly mandating OA dissemination while, at the same time, the growth in costs have led more and more university libraries to cancel some subscriptions.

In this context, we studied the effect of funding on the publication modality and the citations received in more than 128 thousand research articles, of which 31% were funded. These research articles come from 40 randomly selected subject categories in the year 2016, and the citations received from the period 2016-2020 in the Scopus database.

As main conclusion, we found that funded research articles are generally more cited than unfunded ones, but the open access citation advantage in relation to the paywalled modality is higher for the unfunded articles. This open access citation advantage is strong both across fields and over time, and come mainly from hybrid gold modality and the author self-archiving in open access repositories (green).

To contextualize the results we can mention that most of the research articles in the Scopus database in 2016 were unfunded (68%), and only one out of three articles were funded (32%). Moreover, most of them were paywalled, two out of three unfunded articles (67%) and half of the funded articles (51%). The prevalence of unfunded articles rises to 73% in the paywalled modality.

In the funded group there is a greater concern about offering open access to publications. Thus, the use of open access repositories (green) is more widespread within the funded group (41%), compared to the group without funding (24%). The prevalence of the gold modality is quite similar because some gold journals are also free of charge for the authors. However, the prevalence of the hybrid gold modality within the group with funding is double that of the group without funding, motivated because the authors must pay the publication costs under the hybrid gold modality.

There are very important differences among subject categories, both in the prevalence of funding and in the open access modality. Thus, while in the humanities the prevalence of funding is around 5%, in life sciences it exceeds 50% in some cases. Furthermore, while in some social sciences and humanities the prevalence of open access is below 20%, it reaches over 70% in some scientific disciplines.

Interestingly, open articles published in hybrid journals were considerably more cited than those published in open access journals. Thus, articles under the hybrid gold modality are cited on average twice as those in the gold modality. Moreover, 75% of the articles published during 2016 in open access journals (gold) received an average number of citations less than that received by the 25% least cited of open access articles in hybrid journals (hybrid gold). This is the case regardless of funding, so this evidence is strong. However, it should be noted that gold journals are younger than hybrids and in most cases do not have (at the moment) the prestige of the latter.

Within the same publication modality, we found that funded articles generally obtain 50% more citations than unfunded ones. The most cited modality is the hybrid gold and the least cited is the gold, well below even the paywalled. Moreover, the use of open access repositories considerably increases the citations received, especially for those articles without funding. Thus, the articles in open access repositories (green) are 50% more cited than the paywalled ones. This evidence is remarkable and does not depend on funding.

The OA citation advantage is clear for all open access modalities, with the exception of the open access journals (gold) as mentioned. Excluding the gold modality, there is a citation advantage in more than 75% of the subject categories. Furthermore, the citation advantage of open access is considerably greater among unfunded articles. This result is strong both across fields and over time.

The highest advantage reaches in hybrid gold, with 79% and 124% for funded and unfunded, respectively. Half of the categories analyzed present hybrid gold citation advantages greater than 69% for funded and 80% for unfunded articles. In green modality, the average OA citation advantage for funded and unfunded articles is 45% and 81%, respectively. Moreover, half of the categories present green citation advantages greater than 35% for funded and 50% for unfunded. In the only green modality, the average OA citation advantage is 47% for funded and 72% for unfunded articles, although half of the categories present advantages greater than 22% and 61%, respectively.

Furthermore, we found that the OA citation advantage is consistent across subject categories and held in time. In the gold modality, and regardless of funding, although there is a clear citation disadvantage with respect to the paywalled option, this disadvantage decreases over time. However, in the hybrid gold and bronze modalities, where there is a clear citation advantage in relation to the paywalled, this advantage does not depend on time. Finally, the open access modalities in which the trend is more stable are green and only green.

There are some considerations in this regard. Some of the citation advantage is likely due to more access allows more people to read and hence cite articles they otherwise would not. However, causation is difficult to establish and there are many possible bias. Several factors can affect the observed differences in citation rates. Selection bias can be one of them. The selection bias postulate (Craig et al., 2007) suggests that authors choose only their most impactful studies to be open access. The current study does not examine the cause of the observed citation advantage, but does find that it exists in a very large sample that is representative of the general research literature.

Table 1. Representativeness of the sample. Research articles in 2016 by funding and publication modality (Source: Scopus)

| Funding | Modality | Sample | | Total Database |
|---|---|---|---|---|
| Funded Articles | Gold | 4,713 | 4.8% | 98,424 |
| | Hybrid Gold | 2,734 | 8.8% | 31,009 |
| | Bronze | 7,134 | 8.2% | 87,443 |
| | Green | 17,729 | 6.8% | 260,903 |
| | Only Green | 7,796 | 7.9% | 98,788 |
| | Paywalled | 17,298 | 5.3% | 324,368 |
| | All* | 39,675 | 6.2% | 640,032 |
| Unfunded Articles | Gold | 11,128 | 5.8% | 190,257 |
| | Hybrid Gold | 2,492 | 7.4% | 33,545 |
| | Bronze | 7,746 | 6.6% | 118,024 |
| | Green | 22,616 | 7.0% | 325,127 |
| | Only Green | 10,127 | 9.2% | 110,187 |
| | Paywalled | 57,495 | 6.4% | 899,990 |
| | All* | 88,988 | 6.6% | 1,352,003 |
| Total | | 128,663 | 6.5% | 1,992,035 |

* All = Gold + Hybrid Gold + Bronze + Only Green + Paywalled

Table 2. Prevalence of the publication modality by funding. Research articles in the sample and database in 2016 (Source: Scopus)

| Funding | Modality | Sample | | Total Database | |
|---|---|---|---|---|---|
| Funded Articles | Gold | 4,713 | 11.9% | 98,424 | 15.4% |
| | Hybrid Gold | 2,734 | 6.9% | 31,009 | 4.8% |
| | Bronze | 7,134 | 18.0% | 87,443 | 13.7% |
| | Green | 17,729 | 44.7% | 260,903 | 40.8% |
| | Only Green | 7,796 | 19.6% | 98,788 | 15.4% |
| | Paywalled | 17,298 | 43.6% | 324,368 | 50.7% |
| | All* | 39,675 | | 640,032 | |
| Unfunded Articles | Gold | 11,128 | 12.5% | 190,257 | 14.1% |
| | Hybrid Gold | 2,492 | 2.8% | 33,545 | 2.5% |
| | Bronze | 7,746 | 8.7% | 118,024 | 8.7% |
| | Green | 22,616 | 25.4% | 325,127 | 24.0% |
| | Only Green | 10,127 | 11.4% | 110,187 | 8.1% |
| | Paywalled | 57,495 | 64.6% | 899,990 | 66.6% |
| | All* | 88,988 | | 1,352,003 | |

* All = Gold + Hybrid Gold + Bronze + Only Green + Paywalled

Table 3. Prevalence of funding by publication modality. Research articles in the sample and database in 2016 (Source: Scopus)

| | **Modality** | **Funded Articles** | | **Unfunded Articles** | | **Total** |
|---|---|---|---|---|---|---|
| Sample | Gold | 4,713 | 29.8% | 11,128 | 70.2% | 15,841 |
| | Hybrid Gold | 2,734 | 52.3% | 2,492 | 47.7% | 5,226 |
| | Bronze | 7,134 | 47.9% | 7,746 | 52.1% | 14,880 |
| | Green | 17,729 | 43.9% | 22,616 | 56.1% | 40,345 |
| | Only Green | 7,796 | 43.5% | 10,127 | 56.5% | 17,923 |
| | Paywalled | 17,298 | 23.1% | 57,495 | 76.9% | 74,793 |
| | All* | 39,675 | 30.8% | 88,988 | 69.2% | 128,663 |
| Total Database | Gold | 98,424 | 34.1% | 190,257 | 65.9% | 288,681 |
| | Hybrid Gold | 31,009 | 48.0% | 33,545 | 52.0% | 64,554 |
| | Bronze | 87,443 | 42.6% | 118,024 | 57.4% | 205,467 |
| | Green | 260,903 | 44.5% | 325,127 | 55.5% | 586,030 |
| | Only Green | 98,788 | 47.3% | 110,187 | 52.7% | 208,975 |
| | Paywalled | 324,368 | 26.5% | 899,990 | 73.5% | 1,224,358 |
| | All* | 640,032 | 32.1% | 1,352,003 | 67.9% | 1,992,035 |

* All = Gold + Hybrid Gold + Bronze + Only Green + Paywalled

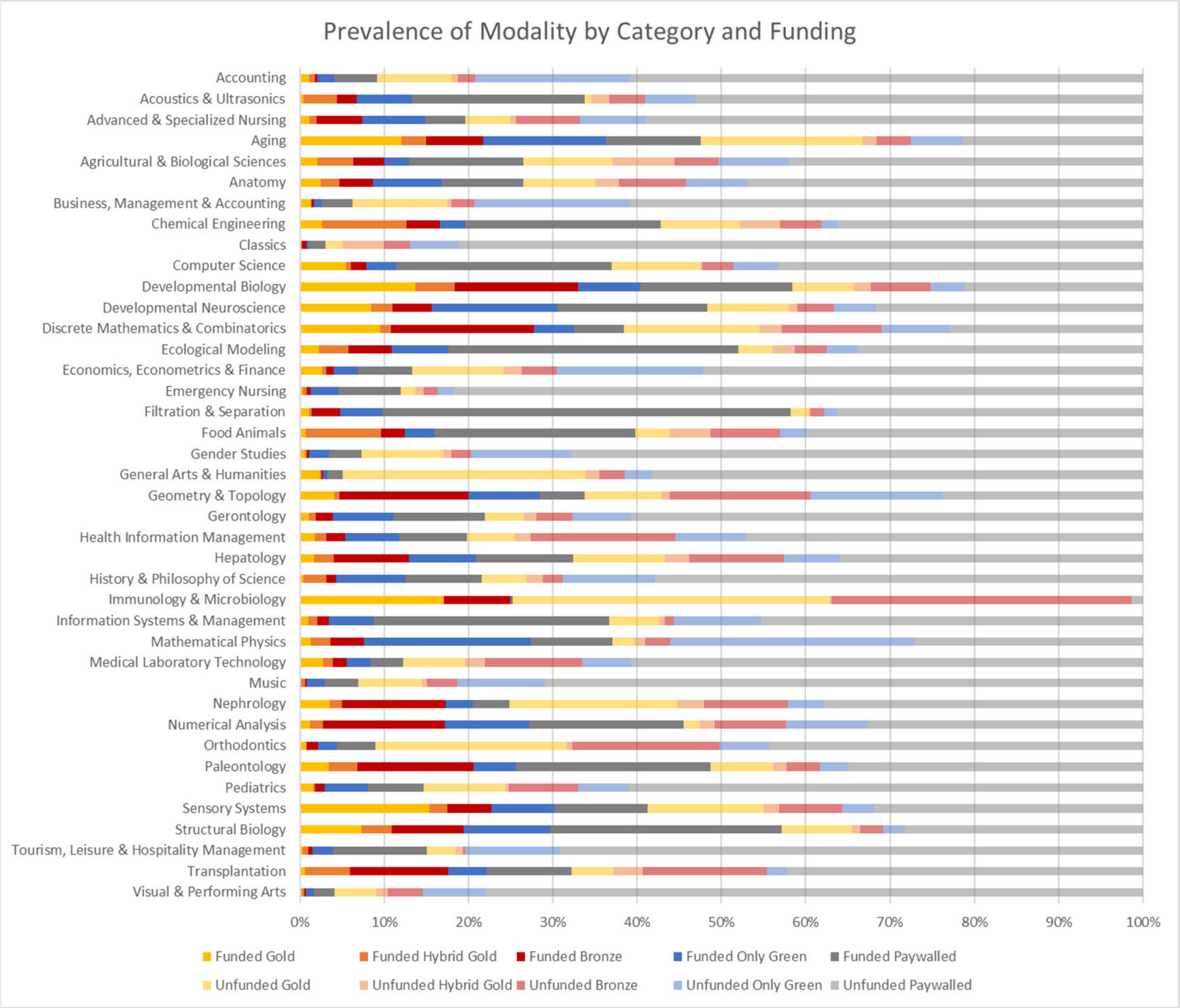

Figure 1. Prevalence of the publication modality by subject category and funding. Research articles in the sample in 2016 (Source: Scopus)

Table 4. Cites per article in 2016-2020 by funding, publication modality and branch of knowledge. Research articles in the sample in 2016 (Source: Scopus)

| | | Health Sciences | | Life Sciences | | Physical Sciences & Engineering | | Social Sciences & Humanities | | Total | |
|---|---|---|---|---|---|---|---|---|---|---|---|
| **Funding** | **Modality** | **Mean** | **Median** | **Mean** | **Median** | **Mean** | **Median** | **Mean** | **Median** | **Mean** | **Median** |
| Funded | Gold | 8.3 | 7.7 | 15.2 | 13.7 | 8.3 | 6.7 | 6.7 | 5.3 | 9.1 | 7.5 |
| | Hybrid Gold | 22.7 | 21.3 | 19.8 | 21.0 | 18.6 | 18.9 | 14.9 | 17.7 | 19.0 | 20.0 |
| | Bronze | 20.5 | 18.8 | 15.5 | 17.4 | 13.2 | 11.3 | 11.5 | 10.4 | 15.3 | 14.1 |
| | Green | 17.5 | 16.3 | 16.9 | 18.0 | 14.2 | 14.2 | 11.1 | 8.1 | 14.8 | 14.5 |
| | Only Green | 15.5 | 13.3 | 15.0 | 14.4 | 15.1 | 13.6 | 12.0 | 10.8 | 14.4 | 13.7 |
| | Paywalled | 11.4 | 10.9 | 12.0 | 12.9 | 12.4 | 11.5 | 9.5 | 8.4 | 11.2 | 10.7 |
| Unfunded | Gold | 5.3 | 4.7 | 11.4 | 11.7 | 6.8 | 6.1 | 3.7 | 3.5 | 6.3 | 5.7 |
| | Hybrid Gold | 16.1 | 15.2 | 15.1 | 13.8 | 13.3 | 8.7 | 9.3 | 8.1 | 13.4 | 12.2 |
| | Bronze | 17.3 | 11.8 | 11.9 | 11.0 | 7.3 | 5.0 | 5.4 | 4.7 | 10.6 | 7.9 |
| | Green | 11.3 | 10.5 | 11.6 | 11.3 | 9.7 | 9.7 | 7.3 | 6.7 | 9.9 | 9.8 |
| | Only Green | 11.4 | 9.6 | 11.1 | 10.6 | 11.3 | 11.2 | 8.5 | 7.8 | 10.5 | 9.5 |
| | Paywalled | 6.4 | 6.1 | 7.2 | 8.3 | 7.1 | 6.0 | 5.9 | 5.5 | 6.6 | 6.1 |

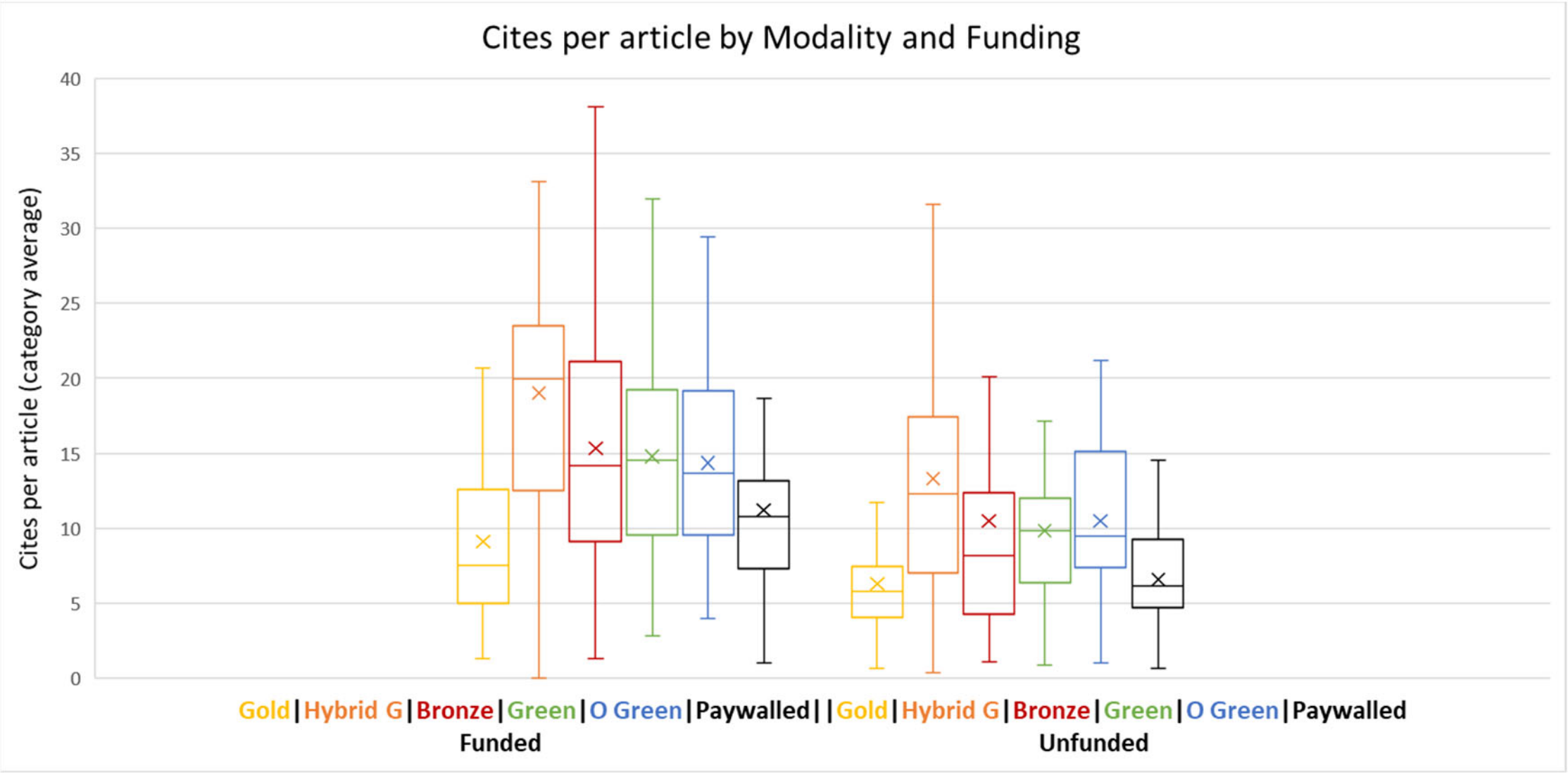

Figure 2. Box and whisker plot (without outliers) for the distribution of cites per article by funding and modality. Research articles in the sample in 2016 and cites in 2016-2020 (Source: Scopus)

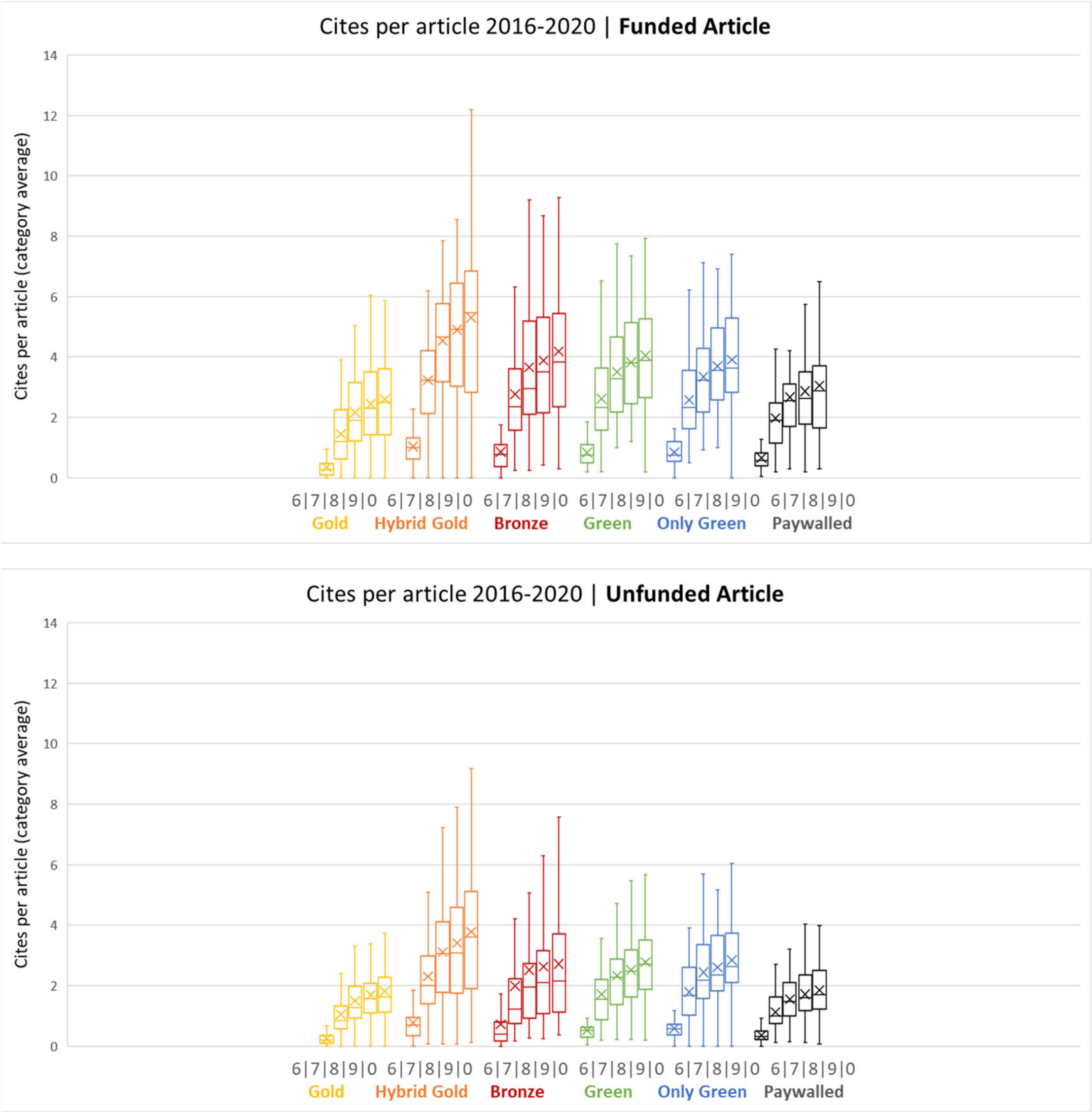

Figure 3. Trend of cites per article by funding and modality. Research articles in the sample in 2016 and cites in 2016-2020 (Source: Scopus)

Table 5. OA citation advantage in 2016-2020 by funding, publication modality and branch of knowledge. Research articles in the sample in 2016 (Source: Scopus)

| | | **Health Sciences** | | **Life Sciences** | | **Physical Sciences & Engineering** | | **Social Sciences & Humanities** | | **Total** | |
|---|---|---|---|---|---|---|---|---|---|---|---|
| **Funding** | **Modality** | **Mean** | **Median** | **Mean** | **Median** | **Mean** | **Median** | **Mean** | **Median** | **Mean** | **Median** |
| Funded | Gold | -51% | -42% | 49% | 21% | -66% | -22% | -88% | -66% | -46% | -36% |
| | Hybrid Gold | 93% | 83% | 76% | 67% | 59% | 60% | 86% | 83% | 79% | 69% |
| | Bronze | 80% | 57% | 40% | 34% | 14% | 3% | 24% | 29% | 41% | 37% |
| | Green | 55% | 41% | 58% | 38% | 27% | 9% | 42% | 49% | 45% | 35% |
| | Only Green | 38% | 29% | 24% | 23% | 37% | 15% | 82% | 37% | 47% | 22% |
| Unfunded | Gold | -22% | -10% | 172% | 39% | 0% | -4% | -63% | -27% | 6% | -6% |
| | Hybrid Gold | 190% | 151% | 152% | 115% | 84% | 56% | 72% | 69% | 124% | 80% |
| | Bronze | 164% | 91% | 188% | 67% | -19% | -13% | 13% | 31% | 81% | 42% |
| | Green | 81% | 63% | 181% | 78% | 42% | 44% | 54% | 27% | 81% | 50% |
| | Only Green | 88% | 71% | 39% | 37% | 62% | 64% | 82% | 55% | 72% | 61% |

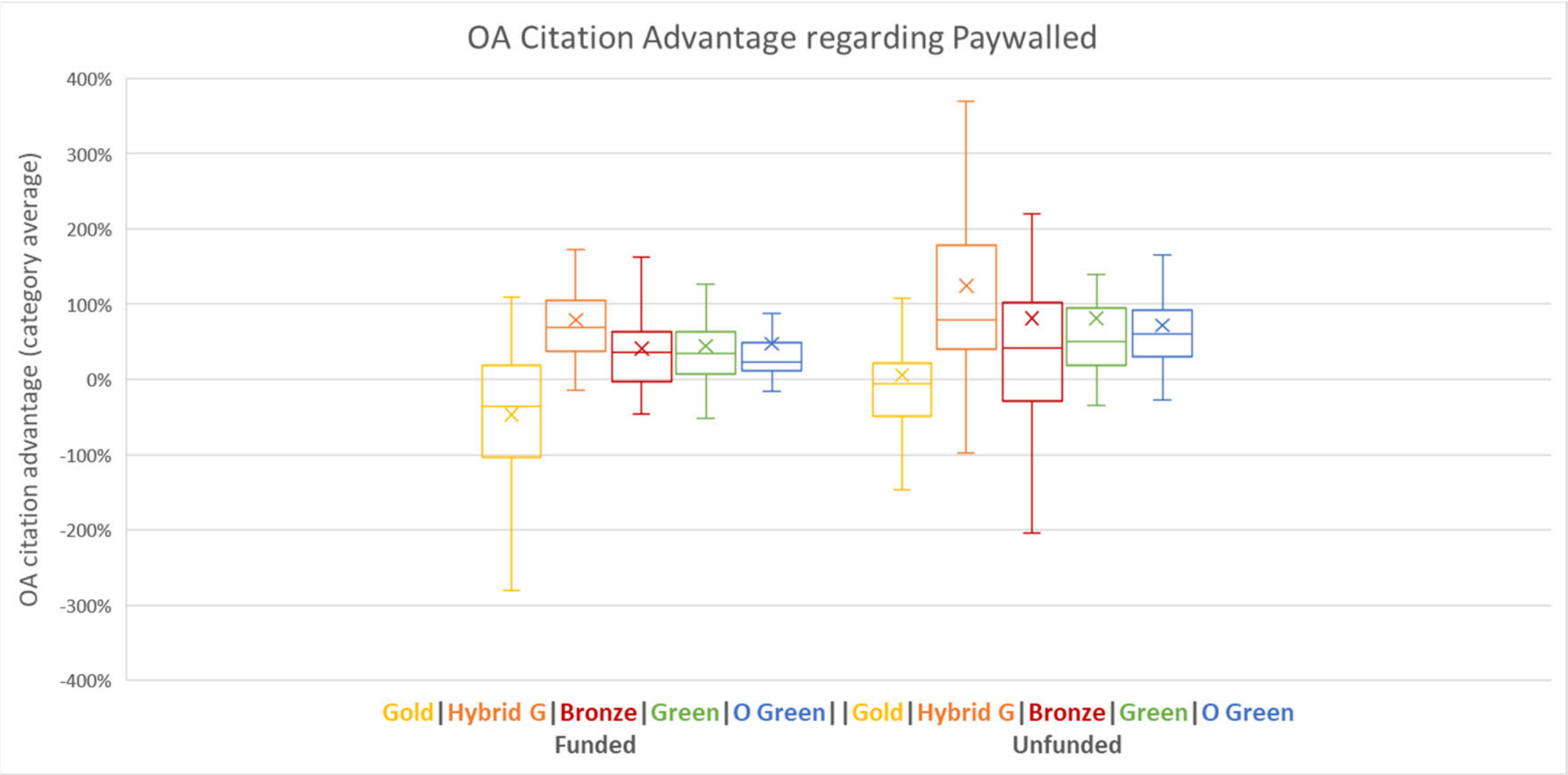

Figure 4. OA citation advantage by funding and modality. Research articles in the sample in 2016 and cites in 2016-2020 (Source: Scopus)

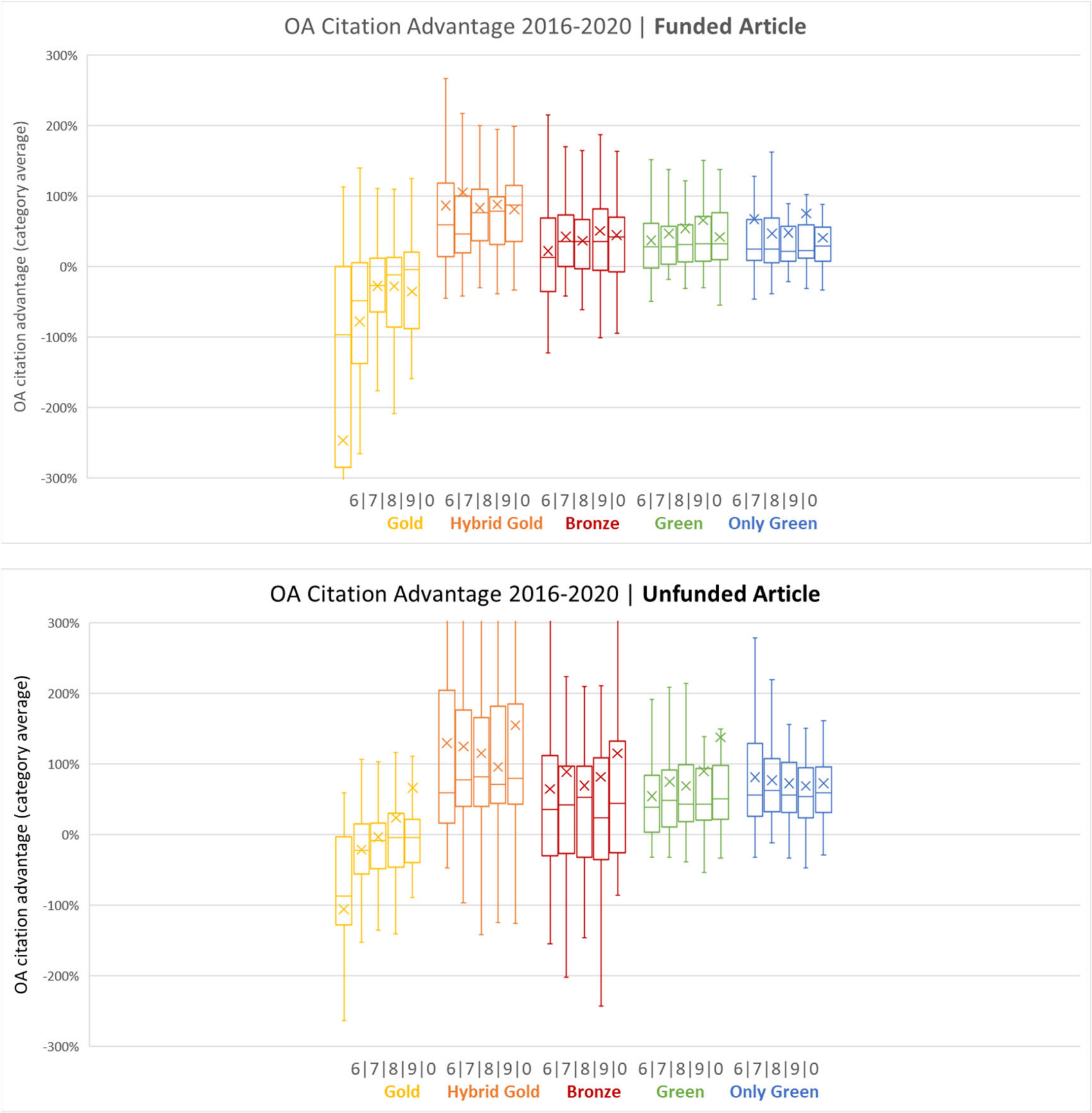

Figure 5. Trend of OA citation advantage by funding and modality. Research articles in the sample in 2016 and cites in 2016-2020 (Source: Scopus)

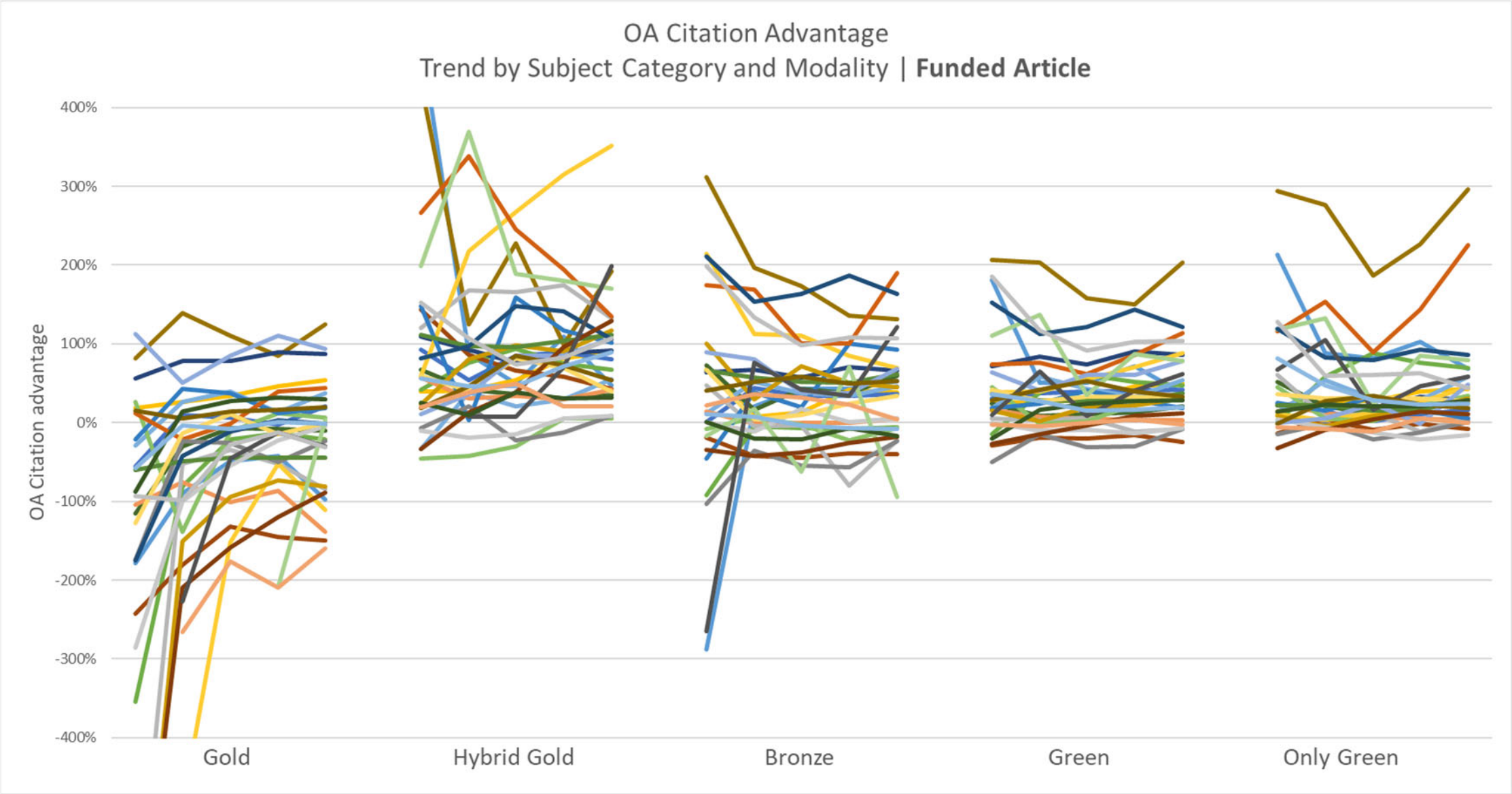

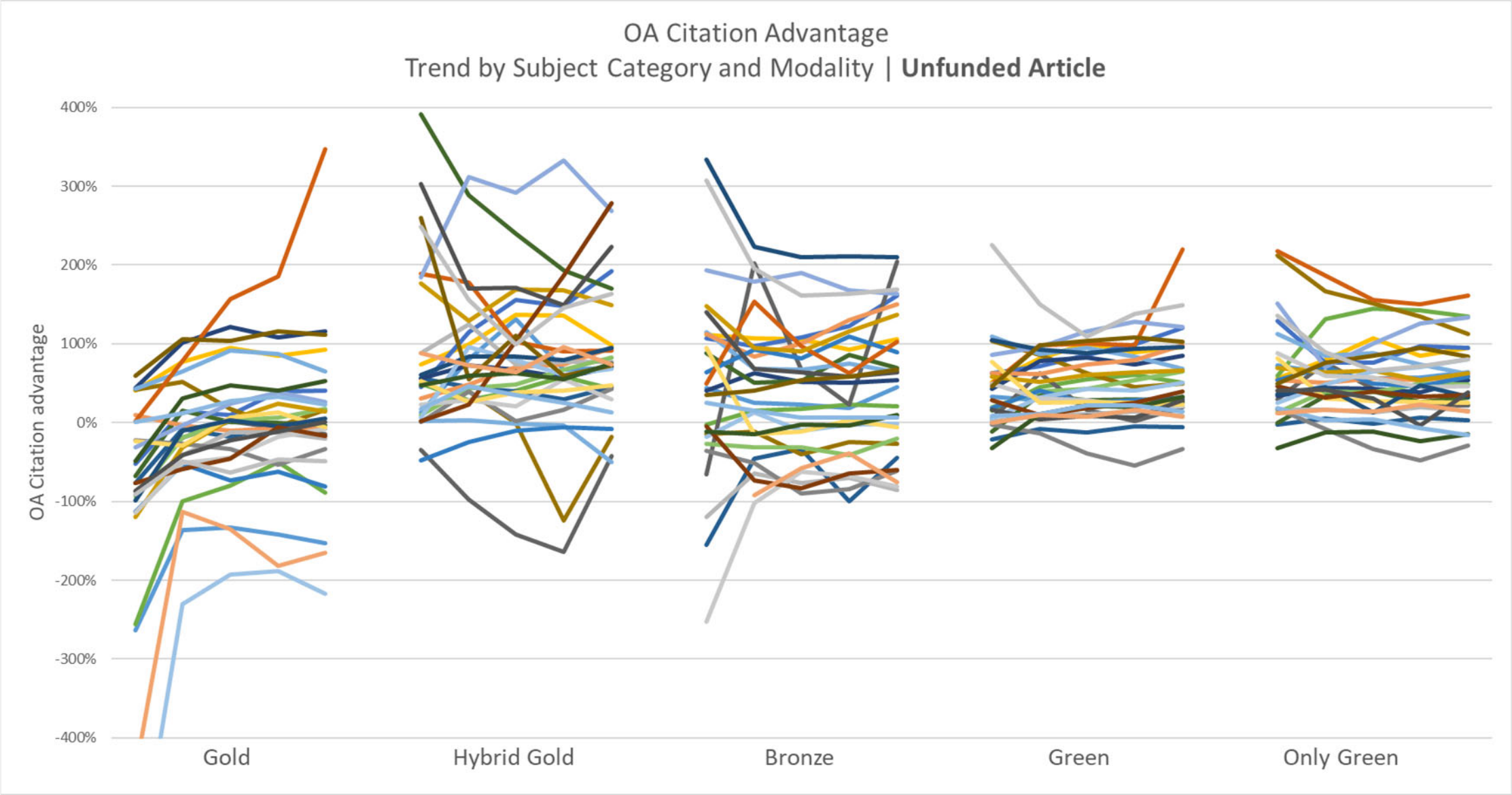

Figure 6. Trend of OA citation advantage by subject category, funding and modality. Research articles in the sample in 2016 and cites in 2016-2020 (Source: Scopus)

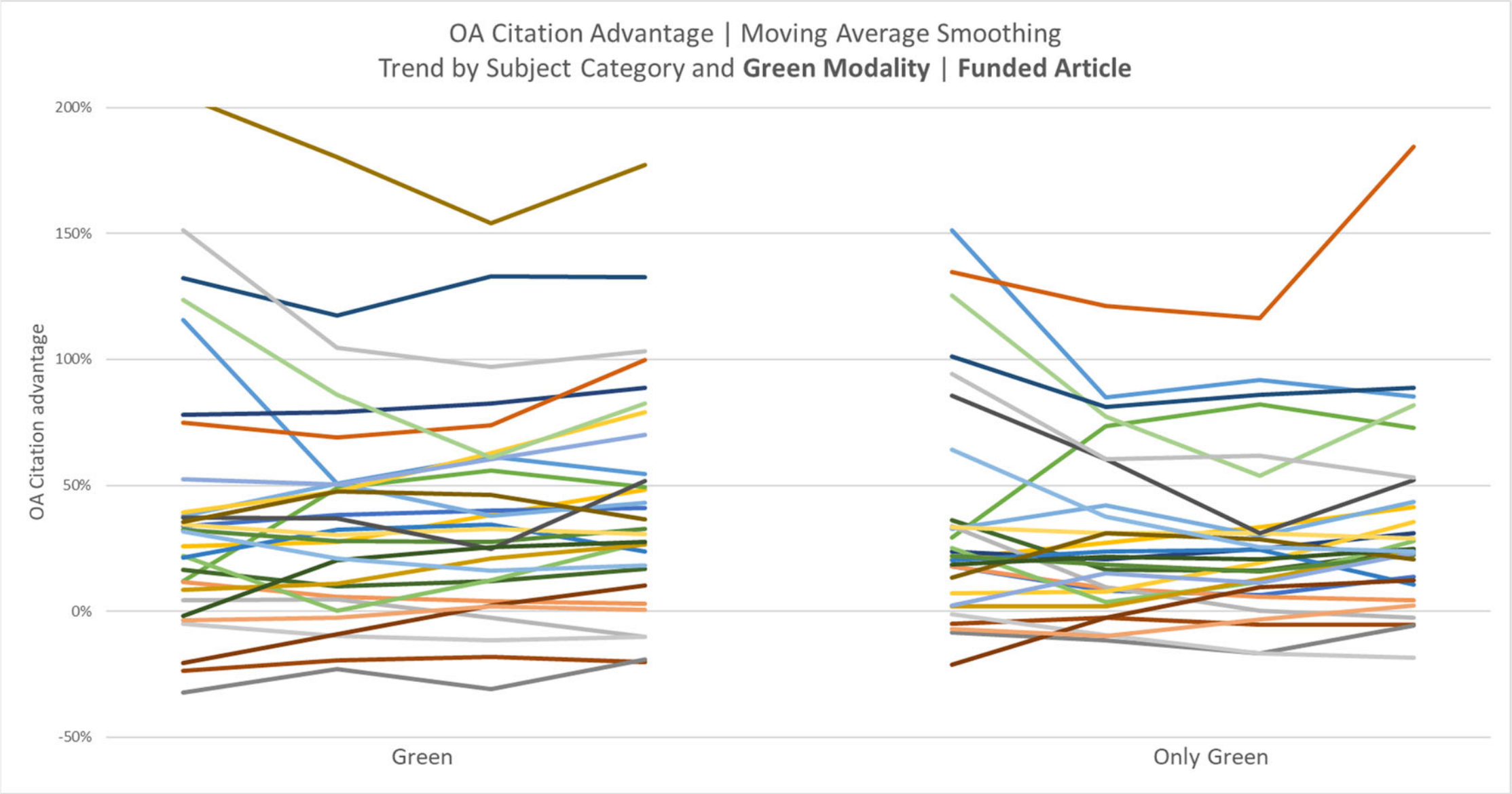

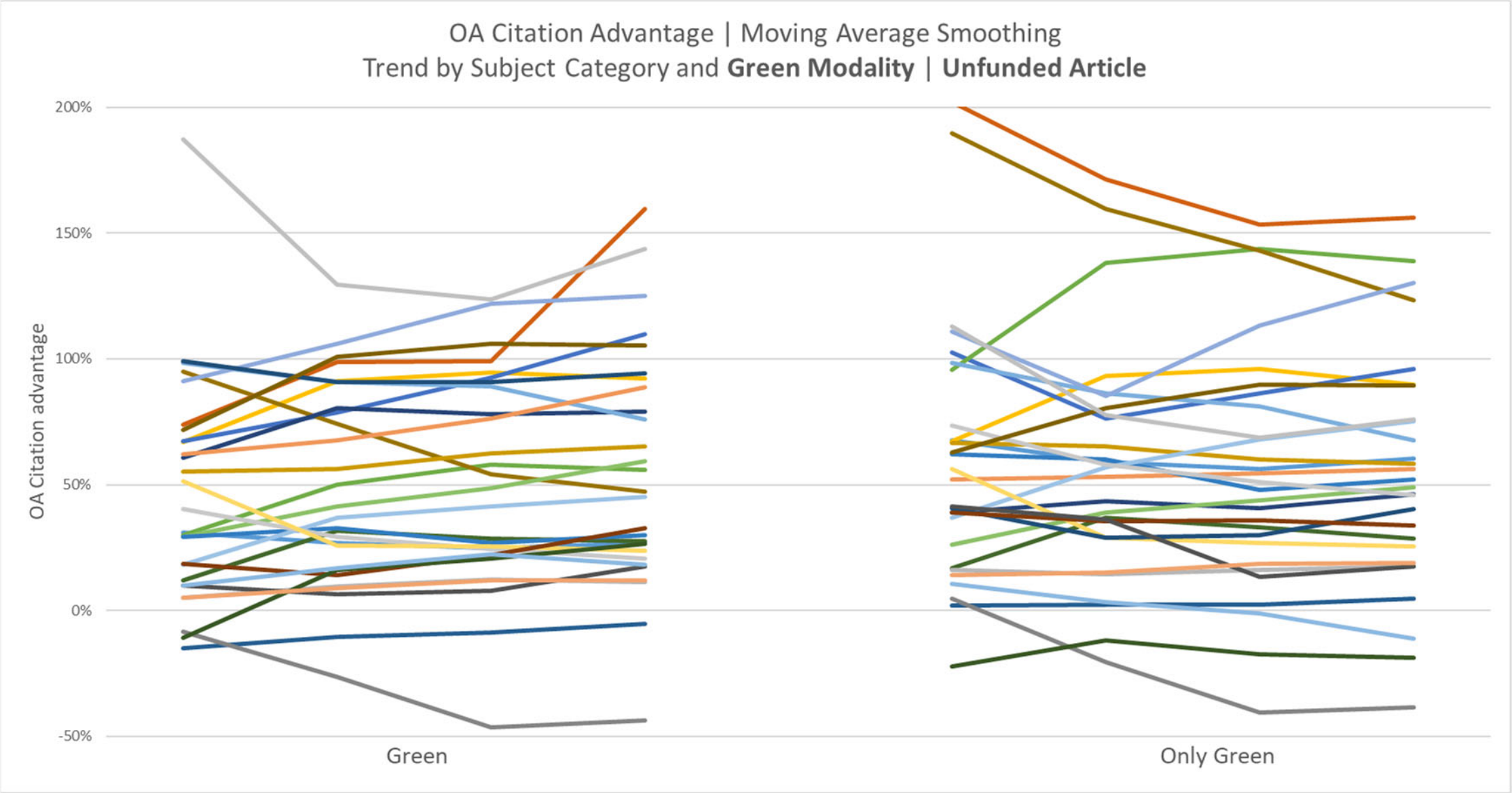

Figure 7. Trend of OA citation advantage by subject category, funding and green modality. Research articles in the sample in 2016 and cites moving average smoothing in 2016-2020 (Source: Scopus)

## References

Aagaard, K., A. Kladakis, & M. W. Nielsen (2020) Concentration or dispersal of research funding? *Quantitative Science Studies*, 1(1), pp. 117–149.

Antelman, K. (2004) Do open-access articles have a greater research impact? *College & Research Libraries*, 65(5), pp. 372–382.

Archambault, E., D. Amyot, P. Deschamps, A. Nicol, F. Provencher, L. Rebout, & G. Roberge (2014) Proportion of open access papers published in peer-reviewed journals at the European and world levels - 1996–2013. Report for the European Commission. Available at https://digitalcommons.unl.edu/scholcom/8/

Björk, B. C. (2004) Open access to scientific publications: an analysis of the barriers to change? *Information Research*, 9(2), pp. 170–170.

Björk, B. C. (2017) Growth of hybrid open access, 2009-2016. *PeerJ*, 5, e3878.

Björk, B. C., & D. Solomon (2012) Open access versus subscription journals: a comparison of scientific impact. *BMC Medicine*, 10(1), pp. 73–73.

Craig, I. D., A. M. Plume, M. E. McVeigh, J. Pringle, & M. Amin (2007) Do open access articles have greater citation impact? A critical review of the literature. *Journal of Informetrics*, 1(3), pp. 239–248.

Davis, P. M. (2011) Open access, readership, citations: a randomized controlled trial of scientific journal publishing. *FASEB Journal*, 25, pp. 2129–2134.

Davis, P. M., B. V. Lewenstein, D. H. Simon, J. G. Booth, & M. J. Connolly (2008) Open access publishing, article downloads, and citations: randomised controlled trial. *British Medical Journal*, 337(7665), pp. 343–345.

Davis, P. M., & W. H. Walters (2011) The impact of free access to the scientific literature: a review of recent research. *Journal of the Medical Library Association*, 99(3), pp. 208–208.

Dorta-González, P., S. M. González-Betancor, & M. I. Dorta-González (2017) Reconsidering the gold open access citation advantage postulate in a multidisciplinary context: an analysis of the subject categories in the Web of Science database 2009-2014. *Scientometrics*, 112(2), pp. 877–901.

Dorta-González, P., & Y. Santana-Jiménez (2018) Prevalence and citation advantage of gold open access in the subject areas of the Scopus database. *Research Evaluation*, 27(1), pp. 1–15.

Dorta-González, P., R. Suárez-Vega, & M. I. Dorta-González (2020) Open access effect on uncitedness: a large-scale study controlling by discipline, source type and visibility. *Scientometrics*, 124(3), pp. 2619–2644.

Eysenbach, G. (2006) Citation advantage of open access articles. *PLoS Biology*, 4(5), e157.

Gargouri, Y., C. Hajjem, V. Larivière, Y. Gingras, L. Carr, T. Brody, et al. (2010) Self-selected or mandated, open access increases citation impact for higher quality research. *PLoS ONE*, 5(10), e13636.

Gaule, P., & N. Maystre (2011) Getting cited: does open access help? *Research Policy*, 40(10), pp. 1332–1338.

González-Betancor, S. M., & P. Dorta-González (2019) Publication modalities 'article in press' and 'open access' in relation to journal average citation. *Scientometrics*, 120(3), pp. 1209–1223.

Gumpenberger, C., M. A. Ovalle-Perandones, & J. Gorraiz (2013) On the impact of gold open access journals. *Scientometrics*, 96(1), pp. 221–238.

Harnad, S., T. Brody, F. Vallières, L. Carr, S. Hitchcock, Y. Gingras, et al. (2004) The access/impact problem and the green and gold roads to open access. *Serials Review*, 30(4), pp. 310–314.

Joint, N. (2009) The Antaeus column: does the "open access" advantage exist? A librarian's perspective. *Library Review*, 58(7), pp. 477–481.

Koler-Povh, T., P. Južnič, & G. Turk (2014) Impact of open access on citation of scholarly publications in the field of civil engineering. *Scientometrics*, 98(1033), pp. 1033–1045.

Kurtz, M. J., G. Eichhorn, A. Accomazzi, C. Grant, M. Demleitner, E. Henneken, et al. (2005) The effect of use and access on citations. *Information Processing & Management*, 41(6), pp. 1395–1402.

Laakso, M., & B. C. Björk (2013) Delayed open access: an overlooked high-impact category of openly available scientific literature. *Journal of the American Society for Information Science and Technology*, 64(7), pp. 1323–1329.

Lawrence, S. (2001) Free online availability substantially increases a paper's impact. *Nature*, 411(6837), pp. 521–521.

McCabe, M., & C. Snyder (2014) Identifying the effect of open access on citations using a panel of science journals. *Economic Inquiry*, 52(4), pp. 1284–1300.

Moed, H. F. (2007) The effect of open access on citation impact: an analysis of ArXiv's condensed matter section. *Journal of the American Society for Information Science and Technology*, 58(13), pp. 2047–2054.

Niyazov, Y., C. Vogel, R. Price, B. Lund, D. Judd, A. Akil, et al. (2016) Open access meets discoverability: citations to articles posted to Academia.edu. *PLoS ONE,* 11(2), e0148257.

Norris, M., C. Oppenheim, & F. Rowland (2008) The citation advantage of open-access articles. *Journal of the American Society for Information Science and Technology*, 59(12), pp. 1963–1972.

Ottaviani, J. (2016) The post-embargo open access citation advantage: it exists (probably), it's modest (usually), and the rich get richer (of course). *PLoS ONE*, 11(8), e0159614.

Piwowar, H., J. Priem, V. Larivière, J. P. Alperin, L. Matthias, B. Norlander, et al. (2018) The state of OA: a large-scale analysis of the prevalence and impact of Open Access articles. *PeerJ*, 6, e4375.

Prosser, D. (2003) Institutional repositories and open access: the future of scholarly communication. *Information Services and Use*, 23(2), pp. 167–170.

Solomon, D. J., M. Laakso, & B. C. Björk (2013) A longitudinal comparison of citation rates and growth among open access journals. *Journal of Informetrics*, 7(3), pp. 642–650.

Tenopir, T., E. Dalton, L. Christian, M. Jones, M. McCabe, M. Smith, & A. Fish (2017) Imagining a gold open access future: attitudes, behaviors, and funding scenarios among authors of academic scholarship. *College & Research Libraries*, 78(6), pp. 824–843.

Wang, X., C. Liu, W. Mao, & Z. Fang (2015) The open access advantage considering citation, article usage and social media attention. *Scientometrics*, 103(2), pp. 555–564.

ANNEX A. Dataset for the sample. Cites per article and OA citation advantage by subject category, funding and publication modality. Research articles in 2016 and citations in 2016-2020 (Source: Scopus)

| Subject Category | Total Articles | Funding | Publication Modality | Articles 2016 | % | Cites 2016 | Cites 2017 | Cites 2018 | Cites 2019 | Cites 2020 | Total Cites | Cites per Article | OA Citation Advantage |
|---|---|---|---|---|---|---|---|---|---|---|---|---|---|
| Accounting | 3758 | Funded | Gold | 43 | 1.1% | 6 | 30 | 61 | 78 | 81 | 256 | 6.0 | -70% |
| | | | Hybrid Gold | 25 | 0.7% | 57 | 62 | 79 | 135 | 137 | 470 | 18.8 | 86% |
| | | | Bronze | 10 | 0.3% | 1 | 16 | 30 | 37 | 57 | 141 | 14.1 | 39% |
| | | | Green | 142 | 3.8% | 155 | 283 | 452 | 630 | 725 | 2245 | 15.8 | 56% |
| | | | Only Green | 78 | 2.1% | 95 | 195 | 298 | 408 | 489 | 1485 | 19.0 | 88% |
| | | | Paywalled | 188 | 5.0% | 73 | 249 | 396 | 486 | 700 | 1904 | 10.1 | |
| | | Unfunded | Gold | 331 | 8.8% | 35 | 155 | 278 | 357 | 437 | 1262 | 3.8 | -146% |
| | | | Hybrid Gold | 30 | 0.8% | 13 | 60 | 136 | 131 | 170 | 510 | 17.0 | 81% |
| | | | Bronze | 76 | 2.0% | 42 | 105 | 182 | 234 | 370 | 933 | 12.3 | 31% |
| | | | Green | 1006 | 26.8% | 515 | 1430 | 2460 | 3259 | 4353 | 12017 | 11.9 | 27% |
| | | | Only Green | 691 | 18.4% | 459 | 1243 | 2107 | 2823 | 3796 | 10428 | 15.1 | 61% |
| | | | Paywalled | 2286 | 60.8% | 878 | 2531 | 4473 | 5955 | 7647 | 21484 | 9.4 | |
| Acoustics & Ultrasonics | 6350 | Funded | Gold | 26 | 0.4% | 20 | 45 | 62 | 96 | 97 | 320 | 12.3 | 1% |
| | | | Hybrid Gold | 255 | 4.0% | 463 | 1401 | 1636 | 1649 | 1617 | 6766 | 26.5 | 117% |
| | | | Bronze | 147 | 2.3% | 60 | 281 | 352 | 408 | 386 | 1487 | 10.1 | -21% |
| | | | Green | 575 | 9.1% | 415 | 1486 | 1889 | 2110 | 1950 | 7850 | 13.7 | 12% |
| | | | Only Green | 421 | 6.6% | 319 | 1122 | 1385 | 1535 | 1401 | 5762 | 13.7 | 12% |
| | | | Paywalled | 1298 | 20.4% | 729 | 2775 | 3977 | 4350 | 4006 | 15837 | 12.2 | |
| | | Unfunded | Gold | 45 | 0.7% | 9 | 57 | 56 | 82 | 59 | 263 | 5.8 | 7% |
| | | | Hybrid Gold | 137 | 2.2% | 253 | 696 | 865 | 849 | 863 | 3526 | 25.7 | 369% |
| | | | Bronze | 271 | 4.3% | 59 | 331 | 559 | 489 | 530 | 1968 | 7.3 | 32% |
| | | | Green | 558 | 8.8% | 340 | 1172 | 1616 | 1694 | 1522 | 6344 | 11.4 | 107% |
| | | | Only Green | 381 | 6.0% | 247 | 893 | 1221 | 1263 | 1122 | 4746 | 12.5 | 127% |
| | | | Paywalled | 3369 | 53.1% | 706 | 2961 | 4860 | 5303 | 4645 | 18475 | 5.5 | |
| Advanced & Specialized Nursing | 3076 | Funded | Gold | 35 | 1.1% | 5 | 22 | 30 | 42 | 50 | 149 | 4.3 | -163% |
| | | | Hybrid Gold | 26 | 0.8% | 29 | 106 | 150 | 158 | 156 | 599 | 23.0 | 106% |
| | | | Bronze | 168 | 5.5% | 409 | 1375 | 1754 | 1720 | 1799 | 7057 | 42.0 | 275% |
| | | | Green | 407 | 13.2% | 545 | 1927 | 2517 | 2627 | 2700 | 10316 | 25.3 | 126% |
| | | | Only Green | 230 | 7.5% | 189 | 818 | 1060 | 1189 | 1208 | 4464 | 19.4 | 73% |
| | | | Paywalled | 146 | 4.7% | 81 | 278 | 409 | 460 | 407 | 1635 | 11.2 | |
| | | Unfunded | Gold | 162 | 5.3% | 10 | 92 | 148 | 194 | 203 | 647 | 4.0 | -3% |
| | | | Hybrid Gold | 21 | 0.7% | 16 | 68 | 88 | 83 | 66 | 321 | 15.3 | 271% |
| | | | Bronze | 233 | 7.6% | 785 | 1801 | 1833 | 1659 | 1764 | 7842 | 33.7 | 717% |
| | | | Green | 447 | 14.5% | 322 | 1037 | 1375 | 1483 | 1604 | 5821 | 13.0 | 216% |
| | | | Only Green | 240 | 7.8% | 166 | 557 | 762 | 834 | 883 | 3202 | 13.3 | 224% |
| | | | Paywalled | 1815 | 59.0% | 383 | 1318 | 1776 | 1935 | 2064 | 7476 | 4.1 | |
| Aging | 2863 | Funded | Gold | 344 | 12.0% | 304 | 1292 | 1722 | 1773 | 2021 | 7112 | 20.7 | 40% |
| | | | Hybrid Gold | 85 | 3.0% | 88 | 353 | 490 | 544 | 617 | 2092 | 24.6 | 67% |
| | | | Bronze | 193 | 6.7% | 240 | 612 | 814 | 965 | 1015 | 3646 | 18.9 | 28% |
| | | | Green | 1010 | 35.3% | 949 | 3739 | 4911 | 5196 | 5765 | 20560 | 20.4 | 38% |
| | | | Only Green | 418 | 14.6% | 355 | 1581 | 1985 | 2054 | 2273 | 8248 | 19.7 | 34% |
| | | | Paywalled | 320 | 11.2% | 238 | 948 | 1198 | 1125 | 1218 | 4727 | 14.8 | |
| | | Unfunded | Gold | 552 | 19.3% | 370 | 1524 | 2056 | 2118 | 2442 | 8510 | 15.4 | 85% |
| | | | Hybrid Gold | 47 | 1.6% | 39 | 146 | 213 | 231 | 214 | 843 | 17.9 | 115% |
| | | | Bronze | 116 | 4.1% | 117 | 375 | 459 | 466 | 549 | 1966 | 16.9 | 103% |
| | | | Green | 806 | 28.2% | 583 | 2289 | 3096 | 3171 | 3620 | 12759 | 15.8 | 90% |
| | | | Only Green | 175 | 6.1% | 130 | 488 | 696 | 672 | 783 | 2769 | 15.8 | 90% |
| | | | Paywalled | 613 | 21.4% | 292 | 956 | 1175 | 1275 | 1408 | 5106 | 8.3 | |
| Agricultural & Biological Sciences | 3417 | Funded | Gold | 71 | 2.1% | 31 | 163 | 236 | 237 | 307 | 974 | 13.7 | 6% |
| | | | Hybrid Gold | 146 | 4.3% | 191 | 473 | 842 | 947 | 945 | 3398 | 23.3 | 80% |
| | | | Bronze | 127 | 3.7% | 87 | 386 | 531 | 583 | 620 | 2207 | 17.4 | 34% |
| | | | Green | 407 | 11.9% | 362 | 1174 | 1773 | 1965 | 2068 | 7342 | 18.0 | 39% |
| | | | Only Green | 98 | 2.9% | 81 | 236 | 312 | 373 | 411 | 1413 | 14.4 | 11% |
| | | | Paywalled | 464 | 13.6% | 316 | 976 | 1451 | 1594 | 1666 | 6003 | 12.9 | |
| | | Unfunded | Gold | 362 | 10.6% | 52 | 339 | 469 | 650 | 697 | 2207 | 6.1 | 25% |
| | | | Hybrid Gold | 252 | 7.4% | 80 | 445 | 762 | 804 | 1008 | 3099 | 12.3 | 152% |
| | | | Bronze | 179 | 5.2% | 81 | 290 | 440 | 512 | 638 | 1961 | 11.0 | 124% |
| | | | Green | 893 | 26.1% | 318 | 1263 | 1964 | 2289 | 2694 | 8528 | 9.5 | 96% |
| | | | Only Green | 282 | 8.3% | 141 | 410 | 587 | 715 | 753 | 2606 | 9.2 | 89% |
| | | | Paywalled | 1436 | 42.0% | 314 | 1183 | 1699 | 1850 | 1963 | 7009 | 4.9 | |
| Anatomy | 3431 | Funded | Gold | 84 | 2.4% | 23 | 115 | 197 | 222 | 225 | 782 | 9.3 | -38% |
| | | | Hybrid Gold | 76 | 2.2% | 134 | 333 | 419 | 398 | 411 | 1695 | 22.3 | 74% |
| | | | Bronze | 137 | 4.0% | 89 | 326 | 363 | 380 | 419 | 1577 | 11.5 | -11% |
| | | | Green | 535 | 15.6% | 583 | 1834 | 2439 | 2429 | 2547 | 9832 | 18.4 | 43% |
| | | | Only Green | 281 | 8.2% | 351 | 1105 | 1508 | 1480 | 1547 | 5991 | 21.3 | 66% |
| | | | Paywalled | 330 | 9.6% | 411 | 820 | 939 | 988 | 1069 | 4227 | 12.8 | |
| | | Unfunded | Gold | 294 | 8.6% | 44 | 207 | 280 | 334 | 296 | 1161 | 3.9 | -84% |
| | | | Hybrid Gold | 98 | 2.9% | 78 | 177 | 237 | 264 | 267 | 1023 | 10.4 | 44% |
| | | | Bronze | 272 | 7.9% | 142 | 439 | 545 | 572 | 623 | 2321 | 8.5 | 17% |
| | | | Green | 713 | 20.8% | 437 | 1454 | 1892 | 1957 | 2046 | 7786 | 10.9 | 50% |
| | | | Only Green | 250 | 7.3% | 213 | 816 | 1048 | 1037 | 1119 | 4233 | 16.9 | 133% |
| | | | Paywalled | 1609 | 46.9% | 858 | 2265 | 2757 | 2747 | 3064 | 11691 | 7.3 | |
| Business, Management & Accounting | 2841 | Funded | Gold | 38 | 1.3% | 1 | 21 | 41 | 39 | 43 | 145 | 3.8 | -117% |
| | | | Hybrid Gold | 2 | 0.1% | 2 | 0 | 3 | 2 | 3 | 10 | 5.0 | -65% |
| | | | Bronze | 8 | 0.3% | 0 | 2 | 2 | 8 | 14 | 26 | 3.3 | -154% |
| | | | Green | 63 | 2.2% | 12 | 39 | 73 | 101 | 117 | 342 | 5.4 | -52% |
| | | | Only Green | 25 | 0.9% | 10 | 23 | 41 | 61 | 78 | 213 | 8.5 | 3% |
| | | | Paywalled | 104 | 3.7% | 61 | 133 | 158 | 209 | 299 | 860 | 8.3 | |
| | | Unfunded | Gold | 320 | 11.3% | 54 | 285 | 441 | 675 | 757 | 2212 | 6.9 | -9% |
| | | | Hybrid Gold | 15 | 0.5% | 7 | 21 | 34 | 42 | 55 | 159 | 10.6 | 40% |
| | | | Bronze | 77 | 2.7% | 9 | 51 | 94 | 83 | 135 | 372 | 4.8 | -57% |
| | | | Green | 819 | 28.8% | 201 | 729 | 1177 | 1691 | 1945 | 5743 | 7.0 | -8% |
| | | | Only Green | 523 | 18.4% | 153 | 533 | 838 | 1199 | 1364 | 4087 | 7.8 | 3% |
| | | | Paywalled | 1729 | 60.9% | 516 | 1666 | 2808 | 3730 | 4366 | 13086 | 7.6 | |
| Chemical Engineering | 2016 | Funded | Gold | 53 | 2.6% | 13 | 63 | 96 | 95 | 103 | 370 | 7.0 | -152% |
| | | | Hybrid Gold | 201 | 10.0% | 410 | 1246 | 1401 | 1399 | 1389 | 5845 | 29.1 | 65% |
| | | | Bronze | 81 | 4.0% | 57 | 190 | 235 | 256 | 280 | 1018 | 12.6 | -40% |
| | | | Green | 141 | 7.0% | 92 | 396 | 494 | 534 | 550 | 2066 | 14.7 | -20% |
| | | | Only Green | 61 | 3.0% | 45 | 211 | 236 | 263 | 273 | 1028 | 16.9 | -5% |
| | | | Paywalled | 466 | 23.1% | 391 | 1553 | 1959 | 2048 | 2260 | 8211 | 17.6 | |
| | | Unfunded | Gold | 191 | 9.5% | 42 | 171 | 303 | 323 | 380 | 1219 | 6.4 | -51% |
| | | | Hybrid Gold | 96 | 4.8% | 199 | 602 | 740 | 709 | 776 | 3026 | 31.5 | 227% |
| | | | Bronze | 99 | 4.9% | 8 | 83 | 90 | 73 | 60 | 314 | 3.2 | -204% |
| | | | Green | 228 | 11.3% | 98 | 348 | 514 | 544 | 612 | 2116 | 9.3 | -4% |
| | | | Only Green | 37 | 1.8% | 43 | 126 | 143 | 150 | 152 | 614 | 16.6 | 72% |
| | | | Paywalled | 731 | 36.3% | 308 | 1199 | 1639 | 1850 | 2056 | 7052 | 9.6 | |
| Classics | 1050 | Funded | Gold | 0 | 0.0% | | | | | | | | |
| | | | Hybrid Gold | 2 | 0.2% | 0 | 0 | 0 | 0 | 0 | 0 | 0.0 | |
| | | | Bronze | 7 | 0.7% | 0 | 2 | 2 | 3 | 2 | 9 | 1.3 | 29% |
| | | | Green | 5 | 0.5% | 1 | 1 | 5 | 6 | 1 | 14 | 2.8 | 180% |
| | | | Only Green | 2 | 0.2% | 1 | 1 | 4 | 6 | 0 | 12 | 6.0 | 500% |
| | | | Paywalled | 21 | 2.0% | 1 | 4 | 6 | 4 | 6 | 21 | 1.0 | |
| | | Unfunded | Gold | 21 | 2.0% | 0 | 2 | 4 | 4 | 3 | 13 | 0.6 | -16% |

| | | | | | | | | | | | | |
|---|---|---|---|---|---|---|---|---|---|---|---|---|
| | | | Hybrid Gold | 52 | 5.0% | 2 | 3 | 4 | 4 | 6 | 19 | 0.4 | -97% |
| | | | Bronze | 32 | 3.0% | 1 | 11 | 10 | 8 | 16 | 46 | 1.4 | 100% |
| | | | Green | 92 | 8.8% | 5 | 17 | 20 | 19 | 18 | 79 | 0.9 | 19% |
| | | | Only Green | 60 | 5.7% | 4 | 12 | 16 | 15 | 12 | 59 | 1.0 | 37% |
| | | | Paywalled | 853 | 81.2% | 44 | 97 | 159 | 173 | 140 | 613 | 0.7 | |
| Computer Science | 2241 | Funded | Gold | 123 | 5.5% | 58 | 303 | 388 | 395 | 394 | 1538 | 12.5 | 110% |
| | | | Hybrid Gold | 13 | 0.6% | 18 | 30 | 64 | 45 | 54 | 211 | 16.2 | 173% |
| | | | Bronze | 41 | 1.8% | 44 | 125 | 168 | 168 | 135 | 640 | 15.6 | 162% |
| | | | Green | 208 | 9.3% | 166 | 648 | 805 | 908 | 899 | 3426 | 16.5 | 177% |
| | | | Only Green | 79 | 3.5% | 81 | 306 | 340 | 449 | 445 | 1621 | 20.5 | 245% |
| | | | Paywalled | 572 | 25.5% | 149 | 588 | 859 | 996 | 814 | 3406 | 6.0 | |
| | | Unfunded | Gold | 237 | 10.6% | 84 | 360 | 447 | 407 | 453 | 1751 | 7.4 | 17% |
| | | | Hybrid Gold | 5 | 0.2% | 0 | 7 | 8 | 4 | 7 | 26 | 5.2 | -21% |
| | | | Bronze | 82 | 3.7% | 18 | 74 | 94 | 118 | 106 | 410 | 5.0 | -26% |
| | | | Green | 289 | 12.9% | 148 | 539 | 754 | 754 | 710 | 2905 | 10.1 | 60% |
| | | | Only Green | 120 | 5.4% | 94 | 322 | 484 | 505 | 419 | 1824 | 15.2 | 141% |
| | | | Paywalled | 969 | 43.2% | 243 | 973 | 1553 | 1741 | 1594 | 6104 | 6.3 | |
| Developmental Biology | 6867 | Funded | Gold | 940 | 13.7% | 893 | 3674 | 4738 | 4763 | 5017 | 19085 | 20.3 | 82% |
| | | | Hybrid Gold | 319 | 4.6% | 405 | 1351 | 1592 | 1600 | 1743 | 6691 | 21.0 | 88% |
| | | | Bronze | 1010 | 14.7% | 1005 | 3697 | 4475 | 4600 | 4788 | 18565 | 18.4 | 65% |
| | | | Green | 2328 | 33.9% | 2432 | 9403 | 11450 | 11863 | 12383 | 47531 | 20.4 | 83% |
| | | | Only Green | 502 | 7.3% | 375 | 1371 | 1658 | 1782 | 1847 | 7033 | 14.0 | 26% |
| | | | Paywalled | 1238 | 18.0% | 753 | 2716 | 3500 | 3305 | 3534 | 13808 | 11.2 | |
| | | Unfunded | Gold | 498 | 7.3% | 347 | 1573 | 2244 | 2162 | 2309 | 8635 | 17.3 | 108% |
| | | | Hybrid Gold | 140 | 2.0% | 106 | 380 | 475 | 460 | 511 | 1932 | 13.8 | 66% |
| | | | Bronze | 492 | 7.2% | 337 | 1257 | 1519 | 1546 | 1631 | 6290 | 12.8 | 54% |
| | | | Green | 1178 | 17.2% | 817 | 3297 | 4377 | 4269 | 4677 | 17437 | 14.8 | 78% |
| | | | Only Green | 277 | 4.0% | 180 | 626 | 806 | 801 | 918 | 3331 | 12.0 | 44% |
| | | | Paywalled | 1451 | 21.1% | 705 | 2275 | 2952 | 3031 | 3116 | 12079 | 8.3 | |
| Developmental Neuroscience | 2301 | Funded | Gold | 196 | 8.5% | 72 | 442 | 671 | 673 | 657 | 2515 | 12.8 | -15% |
| | | | Hybrid Gold | 58 | 2.5% | 77 | 242 | 288 | 282 | 291 | 1180 | 20.3 | 38% |
| | | | Bronze | 105 | 4.6% | 144 | 359 | 546 | 535 | 622 | 2206 | 21.0 | 42% |
| | | | Green | 647 | 28.1% | 634 | 2066 | 2619 | 2713 | 2888 | 10920 | 16.9 | 14% |
| | | | Only Green | 345 | 15.0% | 414 | 1223 | 1415 | 1531 | 1639 | 6222 | 18.0 | 22% |
| | | | Paywalled | 408 | 17.7% | 323 | 1197 | 1489 | 1514 | 1511 | 6034 | 14.8 | |
| | | Unfunded | Gold | 224 | 9.7% | 85 | 427 | 482 | 514 | 543 | 2051 | 9.2 | -1% |
| | | | Hybrid Gold | 22 | 1.0% | 69 | 141 | 159 | 148 | 150 | 667 | 30.3 | 228% |
| | | | Bronze | 100 | 4.3% | 120 | 249 | 325 | 426 | 426 | 1546 | 15.5 | 67% |
| | | | Green | 358 | 15.6% | 205 | 801 | 977 | 1061 | 1136 | 4180 | 11.7 | 26% |
| | | | Only Green | 113 | 4.9% | 72 | 250 | 336 | 327 | 374 | 1359 | 12.0 | 30% |
| | | | Paywalled | 730 | 31.7% | 466 | 1206 | 1553 | 1676 | 1841 | 6742 | 9.2 | |
| Discrete Mathematics & Combinatorics | 5098 | Funded | Gold | 486 | 9.5% | 127 | 418 | 630 | 591 | 663 | 2429 | 5.0 | 24% |
| | | | Hybrid Gold | 63 | 1.2% | 16 | 59 | 71 | 88 | 97 | 331 | 5.3 | 31% |
| | | | Bronze | 868 | 17.0% | 329 | 893 | 1109 | 1249 | 1260 | 4840 | 5.6 | 39% |
| | | | Green | 908 | 17.8% | 362 | 974 | 1209 | 1296 | 1385 | 5226 | 5.8 | 43% |
| | | | Only Green | 240 | 4.7% | 91 | 250 | 292 | 333 | 377 | 1343 | 5.6 | 39% |
| | | | Paywalled | 302 | 5.9% | 102 | 205 | 281 | 325 | 300 | 1213 | 4.0 | |
| | | Unfunded | Gold | 819 | 16.1% | 170 | 605 | 870 | 929 | 945 | 3519 | 4.3 | 75% |
| | | | Hybrid Gold | 135 | 2.6% | 20 | 62 | 74 | 79 | 63 | 298 | 2.2 | -11% |
| | | | Bronze | 605 | 11.9% | 189 | 456 | 562 | 644 | 696 | 2547 | 4.2 | 71% |
| | | | Green | 1063 | 20.9% | 323 | 889 | 1149 | 1186 | 1254 | 4801 | 4.5 | 84% |
| | | | Only Green | 409 | 8.0% | 126 | 337 | 428 | 433 | 462 | 1786 | 4.4 | 78% |
| | | | Paywalled | 1171 | 23.0% | 170 | 523 | 651 | 712 | 821 | 2877 | 2.5 | |
| Ecological Modeling | 3256 | Funded | Gold | 73 | 2.2% | 35 | 169 | 213 | 258 | 213 | 888 | 12.2 | -101% |
| | | | Hybrid Gold | 116 | 3.6% | 179 | 614 | 911 | 992 | 1143 | 3839 | 33.1 | 35% |
| | | | Bronze | 165 | 5.1% | 184 | 659 | 966 | 1081 | 1216 | 4106 | 24.9 | 2% |
| | | | Green | 402 | 12.3% | 458 | 1747 | 2464 | 2734 | 2880 | 10283 | 25.6 | 5% |
| | | | Only Green | 218 | 6.7% | 266 | 986 | 1363 | 1510 | 1574 | 5699 | 26.1 | 7% |
| | | | Paywalled | 1122 | 34.5% | 1100 | 4553 | 6568 | 7403 | 7826 | 27450 | 24.5 | |
| | | Unfunded | Gold | 132 | 4.1% | 76 | 260 | 326 | 376 | 415 | 1453 | 11.0 | -6% |
| | | | Hybrid Gold | 84 | 2.6% | 58 | 253 | 389 | 445 | 551 | 1696 | 20.2 | 73% |
| | | | Bronze | 123 | 3.8% | 138 | 460 | 669 | 865 | 1034 | 3166 | 25.7 | 120% |
| | | | Green | 297 | 9.1% | 255 | 976 | 1400 | 1626 | 1981 | 6238 | 21.0 | 79% |
| | | | Only Green | 122 | 3.7% | 99 | 373 | 517 | 572 | 654 | 2215 | 18.2 | 55% |
| | | | Paywalled | 1101 | 33.8% | 581 | 2238 | 2993 | 3366 | 3705 | 12883 | 11.7 | |
| Economics, Econometrics & Finance | 2643 | Funded | Gold | 70 | 2.6% | 3 | 55 | 95 | 110 | 99 | 362 | 5.2 | -62% |
| | | | Hybrid Gold | 14 | 0.5% | 10 | 45 | 68 | 92 | 85 | 300 | 21.4 | 156% |
| | | | Bronze | 21 | 0.8% | 10 | 24 | 37 | 28 | 45 | 144 | 6.9 | -22% |
| | | | Green | 161 | 6.1% | 55 | 199 | 314 | 345 | 388 | 1301 | 8.1 | -3% |
| | | | Only Green | 77 | 2.9% | 40 | 99 | 158 | 165 | 214 | 676 | 8.8 | 5% |
| | | | Paywalled | 170 | 6.4% | 55 | 204 | 311 | 407 | 444 | 1421 | 8.4 | |
| | | Unfunded | Gold | 288 | 10.9% | 46 | 162 | 316 | 387 | 408 | 1319 | 4.6 | -21% |
| | | | Hybrid Gold | 55 | 2.1% | 31 | 98 | 118 | 147 | 166 | 560 | 10.2 | 84% |
| | | | Bronze | 110 | 4.2% | 15 | 53 | 77 | 98 | 101 | 344 | 3.1 | -77% |
| | | | Green | 758 | 28.7% | 232 | 651 | 1033 | 1307 | 1408 | 4631 | 6.1 | 10% |
| | | | Only Green | 459 | 17.4% | 160 | 423 | 638 | 833 | 899 | 2953 | 6.4 | 16% |
| | | | Paywalled | 1379 | 52.2% | 414 | 1092 | 1702 | 2087 | 2352 | 7647 | 5.5 | |
| Emergency Nursing | 1684 | Funded | Gold | 5 | 0.3% | 0 | 2 | 6 | 10 | 7 | 25 | 5.0 | -139% |
| | | | Hybrid Gold | 9 | 0.5% | 8 | 67 | 100 | 114 | 120 | 409 | 45.4 | 281% |
| | | | Bronze | 8 | 0.5% | 14 | 40 | 51 | 45 | 40 | 190 | 23.8 | 99% |
| | | | Green | 70 | 4.2% | 54 | 230 | 329 | 363 | 388 | 1364 | 19.5 | 63% |
| | | | Only Green | 55 | 3.3% | 34 | 133 | 187 | 211 | 236 | 801 | 14.6 | 22% |
| | | | Paywalled | 124 | 7.4% | 69 | 291 | 375 | 378 | 366 | 1479 | 11.9 | |
| | | Unfunded | Gold | 30 | 1.8% | 2 | 5 | 13 | 23 | 25 | 68 | 2.3 | -40% |
| | | | Hybrid Gold | 16 | 1.0% | 13 | 38 | 29 | 45 | 70 | 195 | 12.2 | 283% |
| | | | Bronze | 28 | 1.7% | 2 | 7 | 12 | 15 | 26 | 62 | 2.2 | -44% |
| | | | Green | 73 | 4.3% | 22 | 62 | 73 | 98 | 127 | 382 | 5.2 | 65% |
| | | | Only Green | 32 | 1.9% | 15 | 48 | 51 | 61 | 82 | 257 | 8.0 | 153% |
| | | | Paywalled | 1377 | 81.8% | 234 | 747 | 1084 | 1107 | 1207 | 4379 | 3.2 | |
| Filtration & Separation | 2219 | Funded | Gold | 25 | 1.1% | 2 | 20 | 37 | 24 | 40 | 123 | 4.9 | -368% |
| | | | Hybrid Gold | 7 | 0.3% | 6 | 25 | 40 | 28 | 42 | 141 | 20.1 | -14% |
| | | | Bronze | 74 | 3.3% | 64 | 366 | 499 | 440 | 408 | 1777 | 24.0 | 4% |
| | | | Green | 151 | 6.8% | 178 | 716 | 977 | 898 | 903 | 3672 | 24.3 | 6% |
| | | | Only Green | 112 | 5.0% | 156 | 609 | 797 | 753 | 743 | 3058 | 27.3 | 19% |
| | | | Paywalled | 1074 | 48.4% | 1118 | 4901 | 6241 | 5972 | 6488 | 24720 | 23.0 | |
| | | Unfunded | Gold | 48 | 2.2% | 11 | 48 | 97 | 84 | 82 | 322 | 6.7 | -84% |
| | | | Hybrid Gold | 4 | 0.2% | 4 | 6 | 8 | 1 | 6 | 25 | 6.3 | -98% |
| | | | Bronze | 37 | 1.7% | 7 | 31 | 50 | 60 | 78 | 226 | 6.1 | -102% |
| | | | Green | 86 | 3.9% | 50 | 165 | 244 | 195 | 215 | 869 | 10.1 | -22% |
| | | | Only Green | 34 | 1.5% | 37 | 103 | 155 | 111 | 128 | 534 | 15.7 | 27% |
| | | | Paywalled | 804 | 36.2% | 563 | 2036 | 2437 | 2387 | 2517 | 9940 | 12.4 | |
| Food Animals | 2736 | Funded | Gold | 21 | 0.8% | 10 | 13 | 40 | 50 | 53 | 166 | 7.9 | -7% |
| | | | Hybrid Gold | 243 | 8.9% | 63 | 252 | 399 | 547 | 608 | 1869 | 7.7 | -10% |
| | | | Bronze | 77 | 2.8% | 27 | 121 | 160 | 134 | 170 | 612 | 7.9 | -7% |
| | | | Green | 155 | 5.7% | 85 | 226 | 335 | 406 | 485 | 1537 | 9.9 | 17% |
| | | | Only Green | 95 | 3.5% | 52 | 148 | 206 | 246 | 302 | 954 | 10.0 | 18% |
| | | | Paywalled | 654 | 23.9% | 247 | 965 | 1394 | 1393 | 1550 | 5549 | 8.5 | |
| | | Unfunded | Gold | 110 | 4.0% | 18 | 104 | 182 | 204 | 252 | 760 | 6.9 | 3% |
| | | | Hybrid Gold | 133 | 4.9% | 46 | 217 | 317 | 387 | 472 | 1439 | 10.8 | 61% |
| | | | Bronze | 224 | 8.2% | 56 | 193 | 274 | 275 | 362 | 1160 | 5.2 | -30% |
| | | | Green | 210 | 7.7% | 79 | 331 | 482 | 561 | 672 | 2125 | 10.1 | 50% |
| | | | Only Green | 87 | 3.2% | 31 | 137 | 193 | 226 | 251 | 838 | 9.6 | 43% |
| | | | Paywalled | 1092 | 39.9% | 345 | 1234 | 1751 | 1897 | 2121 | 7348 | 6.7 | |
| Gender Studies | 3185 | Funded | Gold | 20 | 0.6% | 8 | 39 | 51 | 56 | 78 | 232 | 11.6 | 22% |
| | | | Hybrid Gold | 5 | 0.2% | 6 | 7 | 24 | 27 | 33 | 97 | 19.4 | 105% |
| | | | Bronze | 9 | 0.3% | 3 | 17 | 20 | 45 | 57 | 142 | 15.8 | 67% |
| | | | Green | 105 | 3.3% | 60 | 178 | 273 | 337 | 408 | 1256 | 12.0 | 26% |
| | | | Only Green | 74 | 2.3% | 45 | 116 | 182 | 214 | 256 | 813 | 11.0 | 16% |
| | | | Paywalled | 124 | 3.9% | 60 | 169 | 230 | 309 | 407 | 1175 | 9.5 | |
| | | Unfunded | Gold | 309 | 9.7% | 40 | 158 | 235 | 296 | 338 | 1067 | 3.5 | -71% |
| | | | Hybrid Gold | 32 | 1.0% | 6 | 20 | 38 | 47 | 59 | 170 | 5.3 | -11% |
| | | | Bronze | 75 | 2.4% | 34 | 113 | 179 | 244 | 282 | 852 | 11.4 | 92% |
| | | | Green | 652 | 20.5% | 213 | 713 | 1077 | 1297 | 1707 | 5007 | 7.7 | 30% |
| | | | Only Green | 378 | 11.9% | 161 | 502 | 745 | 859 | 1189 | 3456 | 9.1 | 55% |
| | | | Paywalled | 2159 | 67.8% | 597 | 1685 | 2839 | 3357 | 4289 | 12767 | 5.9 | |
| General Arts & Humanities | 3605 | Funded | Gold | 85 | 2.4% | 26 | 52 | 94 | 126 | 137 | 435 | 5.1 | 19% |
| | | | Hybrid Gold | 8 | 0.2% | 8 | 26 | 31 | 25 | 21 | 111 | 13.9 | 221% |
| | | | Bronze | 8 | 0.2% | 6 | 16 | 18 | 17 | 26 | 83 | 10.4 | 140% |
| | | | Green | 74 | 2.1% | 35 | 97 | 134 | 146 | 177 | 589 | 8.0 | 84% |
| | | | Only Green | 17 | 0.5% | 10 | 32 | 36 | 44 | 62 | 184 | 10.8 | 151% |
| | | | Paywalled | 66 | 1.8% | 18 | 49 | 74 | 70 | 74 | 285 | 4.3 | |
| | | Unfunded | Gold | 1036 | 28.7% | 66 | 273 | 645 | 762 | 1070 | 2816 | 2.7 | 189% |
| | | | Hybrid Gold | 61 | 1.7% | 11 | 25 | 30 | 30 | 27 | 123 | 2.0 | 114% |
| | | | Bronze | 107 | 3.0% | 10 | 40 | 51 | 45 | 50 | 196 | 1.8 | 95% |
| | | | Green | 599 | 16.6% | 56 | 175 | 290 | 307 | 443 | 1271 | 2.1 | 125% |
| | | | Only Green | 116 | 3.2% | 23 | 49 | 72 | 75 | 70 | 289 | 2.5 | 165% |
| | | | Paywalled | 2101 | 58.3% | 131 | 310 | 509 | 542 | 485 | 1977 | 0.9 | |
| Geometry & Topology | 4551 | Funded | Gold | 185 | 4.1% | 32 | 166 | 240 | 216 | 219 | 873 | 4.7 | -36% |
| | | | Hybrid Gold | 27 | 0.6% | 12 | 36 | 36 | 42 | 43 | 169 | 6.3 | -3% |
| | | | Bronze | 697 | 15.3% | 162 | 569 | 735 | 778 | 828 | 3072 | 4.4 | -46% |
| | | | Green | 964 | 21.2% | 304 | 929 | 1198 | 1293 | 1314 | 5038 | 5.2 | -23% |
| | | | Only Green | 387 | 8.5% | 159 | 421 | 520 | 604 | 574 | 2278 | 5.9 | -9% |
| | | | Paywalled | 241 | 5.3% | 114 | 266 | 393 | 422 | 354 | 1549 | 6.4 | |
| | | Unfunded | Gold | 416 | 9.1% | 110 | 294 | 433 | 438 | 456 | 1731 | 4.2 | -37% |
| | | | Hybrid Gold | 43 | 0.9% | 14 | 54 | 61 | 81 | 90 | 300 | 7.0 | 22% |
| | | | Bronze | 760 | 16.7% | 182 | 450 | 557 | 664 | 694 | 2547 | 3.4 | -70% |
| | | | Green | 1445 | 31.8% | 456 | 1136 | 1453 | 1513 | 1590 | 6148 | 4.3 | -34% |
| | | | Only Green | 712 | 15.6% | 271 | 593 | 745 | 776 | 809 | 3194 | 4.5 | -27% |
| | | | Paywalled | 1083 | 23.8% | 351 | 971 | 1510 | 1746 | 1591 | 6169 | 5.7 | |
| Gerontology | 2060 | Funded | Gold | 22 | 1.1% | 2 | 16 | 27 | 33 | 36 | 114 | 5.2 | -103% |
| | | | Hybrid Gold | 17 | 0.8% | 16 | 56 | 80 | 84 | 109 | 345 | 20.3 | 93% |
| | | | Bronze | 41 | 2.0% | 63 | 97 | 167 | 160 | 176 | 663 | 16.2 | 54% |
| | | | Green | 227 | 11.0% | 200 | 423 | 648 | 720 | 880 | 2871 | 12.6 | 20% |
| | | | Only Green | 149 | 7.2% | 124 | 260 | 384 | 453 | 574 | 1795 | 12.0 | 14% |
| | | | Paywalled | 223 | 10.8% | 171 | 407 | 531 | 580 | 659 | 2348 | 10.5 | |
| | | Unfunded | Gold | 97 | 4.7% | 17 | 69 | 130 | 176 | 198 | 590 | 6.1 | 4% |
| | | | Hybrid Gold | 30 | 1.5% | 32 | 64 | 104 | 118 | 134 | 452 | 15.1 | 157% |
| | | | Bronze | 87 | 4.2% | 83 | 158 | 214 | 276 | 369 | 1100 | 12.6 | 115% |
| | | | Green | 301 | 14.6% | 184 | 426 | 624 | 725 | 899 | 2858 | 9.5 | 62% |
| | | | Only Green | 141 | 6.8% | 92 | 216 | 302 | 319 | 410 | 1339 | 9.5 | 62% |
| | | | Paywalled | 1253 | 60.8% | 483 | 1169 | 1615 | 1840 | 2246 | 7353 | 5.9 | |
| Health Information Management | 1594 | Funded | Gold | 29 | 1.8% | 18 | 47 | 65 | 102 | 106 | 338 | 11.7 | -20% |
| | | | Hybrid Gold | 21 | 1.3% | 29 | 52 | 93 | 137 | 177 | 488 | 23.2 | 67% |
| | | | Bronze | 36 | 2.3% | 28 | 140 | 198 | 223 | 185 | 774 | 21.5 | 54% |
| | | | Green | 172 | 10.8% | 195 | 359 | 550 | 655 | 702 | 2461 | 14.3 | 3% |
| | | | Only Green | 101 | 6.3% | 135 | 197 | 278 | 287 | 325 | 1222 | 12.1 | -15% |
| | | | Paywalled | 129 | 8.1% | 158 | 287 | 394 | 482 | 477 | 1798 | 13.9 | |
| | | Unfunded | Gold | 90 | 5.6% | 34 | 115 | 159 | 178 | 204 | 690 | 7.7 | -33% |
| | | | Hybrid Gold | 30 | 1.9% | 22 | 69 | 103 | 123 | 163 | 480 | 16.0 | 57% |
| | | | Bronze | 274 | 17.2% | 1850 | 3912 | 4491 | 3678 | 3125 | 17056 | 62.2 | 509% |
| | | | Green | 365 | 22.9% | 771 | 1822 | 2202 | 1932 | 1723 | 8450 | 23.2 | 126% |
| | | | Only Green | 134 | 8.4% | 231 | 524 | 761 | 692 | 630 | 2838 | 21.2 | 107% |
| | | | Paywalled | 750 | 47.1% | 591 | 1254 | 1727 | 2022 | 2073 | 7667 | 10.2 | |
| Hepatology | 5149 | Funded | Gold | 87 | 1.7% | 97 | 296 | 366 | 345 | 347 | 1451 | 16.7 | -47% |
| | | | Hybrid Gold | 122 | 2.4% | 460 | 1218 | 1456 | 1429 | 1488 | 6051 | 49.6 | 103% |
| | | | Bronze | 455 | 8.8% | 1350 | 3637 | 4190 | 3945 | 4223 | 17345 | 38.1 | 56% |
| | | | Green | 883 | 17.1% | 2138 | 5761 | 6839 | 6497 | 7001 | 28236 | 32.0 | 31% |
| | | | Only Green | 412 | 8.0% | 895 | 2558 | 2884 | 2759 | 3046 | 12142 | 29.5 | 21% |
| | | | Paywalled | 594 | 11.5% | 1058 | 3014 | 3613 | 3417 | 3420 | 14522 | 24.4 | |
| | | Unfunded | Gold | 559 | 10.9% | 263 | 965 | 1191 | 1173 | 1202 | 4794 | 8.6 | -52% |
| | | | Hybrid Gold | 150 | 2.9% | 339 | 870 | 883 | 701 | 548 | 3341 | 22.3 | 71% |
| | | | Bronze | 578 | 11.2% | 990 | 2724 | 3460 | 3642 | 413 | 11229 | 19.4 | 49% |
| | | | Green | 1133 | 22.0% | 934 | 3293 | 3957 | 3834 | 4141 | 16159 | 14.3 | 9% |
| | | | Only Green | 345 | 6.7% | 254 | 1076 | 1314 | 1285 | 1437 | 5366 | 15.6 | 19% |
| | | | Paywalled | 1847 | 35.9% | 1688 | 4999 | 5920 | 5618 | 5872 | 24097 | 13.0 | |
| History & Philosophy of Science | 3653 | Funded | Gold | 14 | 0.4% | 4 | 3 | 7 | 14 | 8 | 36 | 2.6 | -280% |
| | | | Hybrid Gold | 100 | 2.7% | 71 | 268 | 397 | 473 | 559 | 1768 | 17.7 | 81% |
| | | | Bronze | 42 | 1.1% | 26 | 91 | 139 | 139 | 200 | 595 | 14.2 | 45% |
| | | | Green | 418 | 11.4% | 298 | 924 | 1440 | 1595 | 1933 | 6190 | 14.8 | 51% |
| | | | Only Green | 302 | 8.3% | 234 | 642 | 1040 | 1098 | 1343 | 4357 | 14.4 | 47% |
| | | | Paywalled | 328 | 9.0% | 151 | 469 | 718 | 892 | 979 | 3209 | 9.8 | |
| | | Unfunded | Gold | 196 | 5.4% | 9 | 49 | 72 | 84 | 83 | 297 | 1.5 | -217% |
| | | | Hybrid Gold | 71 | 1.9% | 24 | 115 | 137 | 141 | 160 | 577 | 8.1 | 69% |
| | | | Bronze | 85 | 2.3% | 23 | 79 | 83 | 102 | 113 | 400 | 4.7 | -2% |
| | | | Green | 572 | 15.7% | 191 | 623 | 874 | 993 | 1154 | 3835 | 6.7 | 40% |
| | | | Only Green | 399 | 10.9% | 159 | 490 | 709 | 840 | 965 | 3163 | 7.9 | 65% |
| | | | Paywalled | 2116 | 57.9% | 674 | 1749 | 2272 | 2618 | 2840 | 10153 | 4.8 | |
| Immunology & Microbiology | 1171 | Funded | Gold | 199 | 17.0% | 84 | 402 | 515 | 538 | 547 | 2086 | 10.5 | 214% |
| | | | Hybrid Gold | 1 | 0.1% | 0 | 5 | 1 | 2 | 1 | 9 | 9.0 | 170% |
| | | | Bronze | 92 | 7.9% | 24 | 122 | 180 | 167 | 217 | 710 | 7.7 | 132% |
| | | | Green | 289 | 24.7% | 106 | 530 | 695 | 708 | 765 | 2804 | 9.7 | 191% |
| | | | Only Green | 1 | 0.1% | 0 | 1 | 1 | 1 | 1 | 4 | 4.0 | 20% |
| | | | Paywalled | 3 | 0.3% | 3 | 1 | 1 | 1 | 4 | 10 | 3.3 | |
| | | Unfunded | Gold | 440 | 37.6% | 94 | 593 | 757 | 779 | 867 | 3090 | 7.0 | 921% |
| | | | Hybrid Gold | 2 | 0.2% | 0 | 2 | 2 | 1 | 2 | 7 | 3.5 | 409% |
| | | | Bronze | 417 | 35.6% | 100 | 620 | 764 | 789 | 769 | 3042 | 7.3 | 961% |
| | | | Green | 854 | 72.9% | 193 | 1212 | 1521 | 1566 | 1630 | 6122 | 7.2 | 943% |
| | | | Only Green | 0 | 0.0% | | | | | | | | |
| | | | Paywalled | 16 | 1.4% | 0 | 3 | 5 | 2 | 1 | 11 | 0.7 | |
| Information Systems & Management | 4533 | Funded | Gold | 45 | 1.0% | 15 | 96 | 187 | 272 | 316 | 886 | 19.7 | -36% |
| | | | Hybrid Gold | 49 | 1.1% | 57 | 175 | 275 | 379 | 390 | 1276 | 26.0 | -2% |
| | | | Bronze | 61 | 1.3% | 79 | 234 | 459 | 452 | 473 | 1697 | 27.8 | 4% |
| | | | Green | 322 | 7.1% | 414 | 1243 | 1892 | 2088 | 2214 | 7851 | 24.4 | -9% |
| | | | Only Green | 242 | 5.3% | 323 | 969 | 1383 | 1478 | 1531 | 5684 | 23.5 | -14% |
| | | | Paywalled | 1265 | 27.9% | 1625 | 5384 | 8126 | 9343 | 9281 | 33759 | 26.7 | |
| | | Unfunded | Gold | 269 | 5.9% | 77 | 334 | 558 | 805 | 991 | 2765 | 10.3 | -27% |
| | | | Hybrid Gold | 33 | 0.7% | 25 | 80 | 119 | 182 | 170 | 576 | 17.5 | 34% |
| | | | Bronze | 46 | 1.0% | 8 | 43 | 85 | 97 | 101 | 334 | 7.3 | -79% |
| | | | Green | 658 | 14.5% | 606 | 1621 | 2523 | 2827 | 3158 | 10735 | 16.3 | 25% |
| | | | Only Green | 471 | 10.4% | 544 | 1412 | 2214 | 2430 | 2742 | 9342 | 19.8 | 52% |
| | | | Paywalled | 2052 | 45.3% | 1260 | 3866 | 6149 | 7279 | 8174 | 26728 | 13.0 | |
| | 5019 | Funded | Gold | 64 | 1.3% | 14 | 83 | 119 | 92 | 104 | 412 | 6.4 | -8% |
| Mathematical Physics | | | Hybrid Gold | 116 | 2.3% | 92 | 241 | 352 | 326 | 262 | 1273 | 11.0 | 59% |
| | | | Bronze | 204 | 4.1% | 169 | 312 | 376 | 417 | 442 | 1716 | 8.4 | 22% |
| | | | Green | 1313 | 26.2% | 922 | 2473 | 2926 | 2879 | 2751 | 11951 | 9.1 | 31% |
| | | | Only Green | 988 | 19.7% | 669 | 1900 | 2174 | 2131 | 2046 | 8920 | 9.0 | 30% |
| | | | Paywalled | 487 | 9.7% | 242 | 718 | 815 | 806 | 790 | 3371 | 6.9 | |
| | | Unfunded | Gold | 131 | 2.6% | 31 | 108 | 177 | 180 | 149 | 645 | 4.9 | -2% |
| | | | Hybrid Gold | 67 | 1.3% | 30 | 90 | 118 | 115 | 118 | 471 | 7.0 | 39% |
| | | | Bronze | 147 | 2.9% | 84 | 136 | 167 | 184 | 166 | 737 | 5.0 | -1% |
| | | | Green | 1725 | 34.4% | 895 | 2298 | 2753 | 2625 | 2538 | 11109 | 6.4 | 28% |
| | | | Only Green | 1457 | 29.0% | 776 | 2020 | 2354 | 2238 | 2181 | 9569 | 6.6 | 30% |
| | | | Paywalled | 1358 | 27.1% | 398 | 1440 | 1723 | 1653 | 1630 | 6844 | 5.0 | |
| Medical Laboratory Technology | 2050 | Funded | Gold | 57 | 2.8% | 62 | 143 | 186 | 220 | 181 | 792 | 13.9 | 87% |
| | | | Hybrid Gold | 23 | 1.1% | 13 | 52 | 75 | 78 | 78 | 296 | 12.9 | 74% |
| | | | Bronze | 34 | 1.7% | 33 | 102 | 83 | 86 | 94 | 398 | 11.7 | 58% |
| | | | Green | 150 | 7.3% | 126 | 351 | 423 | 444 | 439 | 1783 | 11.9 | 60% |
| | | | Only Green | 57 | 2.8% | 29 | 100 | 125 | 103 | 138 | 495 | 8.7 | 17% |
| | | | Paywalled | 80 | 3.9% | 41 | 133 | 141 | 147 | 131 | 593 | 7.4 | |
| | | Unfunded | Gold | 152 | 7.4% | 33 | 119 | 182 | 198 | 184 | 716 | 4.7 | 19% |
| | | | Hybrid Gold | 47 | 2.3% | 38 | 157 | 178 | 192 | 166 | 731 | 15.6 | 292% |
| | | | Bronze | 237 | 11.6% | 198 | 536 | 664 | 598 | 599 | 2595 | 10.9 | 176% |
| | | | Green | 380 | 18.5% | 201 | 605 | 793 | 817 | 808 | 3224 | 8.5 | 114% |
| | | | Only Green | 119 | 5.8% | 85 | 165 | 230 | 254 | 267 | 1001 | 8.4 | 112% |
| | | | Paywalled | 1244 | 60.7% | 354 | 1010 | 1202 | 1173 | 1193 | 4932 | 4.0 | |
| Music | 1853 | Funded | Gold | 3 | 0.2% | 0 | 0 | 0 | 1 | 3 | 4 | 1.3 | -211% |
| | | | Hybrid Gold | 8 | 0.4% | 7 | 24 | 28 | 23 | 21 | 103 | 12.9 | 210% |
| | | | Bronze | 4 | 0.2% | 1 | 3 | 3 | 7 | 2 | 16 | 4.0 | -4% |
| | | | Green | 52 | 2.8% | 32 | 79 | 85 | 100 | 90 | 386 | 7.4 | 79% |
| | | | Only Green | 39 | 2.1% | 25 | 58 | 58 | 74 | 68 | 283 | 7.3 | 75% |
| | | | Paywalled | 75 | 4.0% | 22 | 48 | 91 | 77 | 73 | 311 | 4.1 | |
| | | Unfunded | Gold | 139 | 7.5% | 5 | 9 | 23 | 29 | 21 | 87 | 0.6 | -251% |
| | | | Hybrid Gold | 12 | 0.6% | 6 | 7 | 8 | 5 | 3 | 29 | 2.4 | 10% |
| | | | Bronze | 66 | 3.6% | 15 | 45 | 49 | 41 | 47 | 197 | 3.0 | 36% |
| | | | Green | 289 | 15.6% | 71 | 200 | 236 | 231 | 270 | 1008 | 3.5 | 59% |
| | | | Only Green | 193 | 10.4% | 57 | 162 | 189 | 185 | 224 | 817 | 4.2 | 93% |
| | | | Paywalled | 1314 | 70.9% | 247 | 506 | 623 | 767 | 742 | 2885 | 2.2 | |
| Nephrology | 4527 | Funded | Gold | 160 | 3.5% | 43 | 281 | 385 | 427 | 451 | 1587 | 9.9 | -14% |
| | | | Hybrid Gold | 65 | 1.4% | 87 | 319 | 431 | 405 | 383 | 1625 | 25.0 | 120% |
| | | | Bronze | 560 | 12.4% | 1288 | 3539 | 3958 | 4145 | 4204 | 17134 | 30.6 | 170% |
| | | | Green | 796 | 17.6% | 1481 | 4225 | 4730 | 5003 | 5039 | 20478 | 25.7 | 127% |
| | | | Only Green | 143 | 3.2% | 232 | 651 | 689 | 709 | 757 | 3038 | 21.2 | 87% |
| | | | Paywalled | 199 | 4.4% | 147 | 496 | 533 | 513 | 568 | 2257 | 11.3 | |
| | | Unfunded | Gold | 897 | 19.8% | 179 | 1042 | 1388 | 1308 | 1471 | 5388 | 6.0 | -4% |
| | | | Hybrid Gold | 146 | 3.2% | 93 | 343 | 404 | 400 | 446 | 1686 | 11.5 | 84% |
| | | | Bronze | 453 | 10.0% | 779 | 1869 | 2111 | 2148 | 2199 | 9106 | 20.1 | 221% |
| | | | Green | 967 | 21.4% | 786 | 2383 | 2745 | 2849 | 2958 | 11721 | 12.1 | 93% |
| | | | Only Green | 187 | 4.1% | 101 | 346 | 318 | 420 | 390 | 1575 | 8.4 | 34% |
| | | | Paywalled | 1717 | 37.9% | 681 | 2192 | 2586 | 2618 | 2686 | 10763 | 6.3 | |
| Numerical Analysis | 4356 | Funded | Gold | 52 | 1.2% | 5 | 39 | 59 | 74 | 84 | 261 | 5.0 | -143% |
| | | | Hybrid Gold | 67 | 1.5% | 39 | 176 | 271 | 410 | 467 | 1363 | 20.3 | 67% |
| | | | Bronze | 631 | 14.5% | 364 | 1027 | 1347 | 1563 | 1628 | 5929 | 9.4 | -30% |
| | | | Green | 832 | 19.1% | 512 | 1689 | 2343 | 2826 | 2824 | 10194 | 12.3 | 0% |
| | | | Only Green | 433 | 9.9% | 254 | 918 | 1326 | 1547 | 1454 | 5499 | 12.7 | 4% |
| | | | Paywalled | 799 | 18.3% | 622 | 1855 | 2344 | 2493 | 2430 | 9744 | 12.2 | |
| | | Unfunded | Gold | 83 | 1.9% | 18 | 68 | 96 | 144 | 133 | 459 | 5.5 | -28% |
| | | | Hybrid Gold | 79 | 1.8% | 31 | 127 | 271 | 415 | 557 | 1401 | 17.7 | 151% |
| | | | Bronze | 366 | 8.4% | 134 | 276 | 338 | 407 | 427 | 1582 | 4.3 | -64% |
| | | | Green | 723 | 16.6% | 355 | 1032 | 1448 | 1664 | 1886 | 6385 | 8.8 | 25% |
| | | | Only Green | 421 | 9.7% | 236 | 723 | 990 | 1023 | 1060 | 4032 | 9.6 | 35% |
| | | | Paywalled | 1425 | 32.7% | 547 | 1858 | 2406 | 2612 | 2658 | 10081 | 7.1 | |
| Orthodontics | 1018 | Funded | Gold | 7 | 0.7% | 0 | 2 | 9 | 11 | 9 | 31 | 4.4 | -47% |
| | | | Hybrid Gold | 1 | 0.1% | 0 | 1 | 2 | 3 | 5 | 11 | 11.0 | 69% |
| | | | Bronze | 14 | 1.4% | 1 | 23 | 37 | 33 | 52 | 146 | 10.4 | 60% |
| | | | Green | 35 | 3.4% | 10 | 54 | 71 | 87 | 95 | 317 | 9.1 | 39% |
| | | | Only Green | 23 | 2.3% | 10 | 44 | 50 | 59 | 61 | 224 | 9.7 | 50% |
| | | | Paywalled | 46 | 4.5% | 12 | 43 | 86 | 81 | 77 | 299 | 6.5 | |
| | | Unfunded | Gold | 231 | 22.7% | 22 | 121 | 229 | 309 | 362 | 1043 | 4.5 | -14% |
| | | | Hybrid Gold | 7 | 0.7% | 5 | 14 | 23 | 26 | 35 | 103 | 14.7 | 185% |
| | | | Bronze | 178 | 17.5% | 76 | 222 | 354 | 421 | 451 | 1524 | 8.6 | 66% |
| | | | Green | 326 | 32.0% | 67 | 251 | 430 | 518 | 648 | 1914 | 5.9 | 14% |
| | | | Only Green | 60 | 5.9% | 15 | 63 | 95 | 86 | 129 | 388 | 6.5 | 25% |
| | | | Paywalled | 451 | 44.3% | 80 | 334 | 546 | 672 | 698 | 2330 | 5.2 | |
| Paleontology | 4846 | Funded | Gold | 168 | 3.5% | 141 | 365 | 480 | 554 | 582 | 2122 | 12.6 | 16% |
| | | | Hybrid Gold | 163 | 3.4% | 141 | 456 | 753 | 796 | 727 | 2873 | 17.6 | 62% |
| | | | Bronze | 666 | 13.7% | 687 | 1927 | 2635 | 2821 | 2950 | 11020 | 16.5 | 52% |
| | | | Green | 745 | 15.4% | 700 | 2022 | 2844 | 2949 | 2882 | 11397 | 15.3 | 41% |
| | | | Only Green | 243 | 5.0% | 176 | 593 | 816 | 843 | 838 | 3266 | 13.4 | 24% |
| | | | Paywalled | 1119 | 23.1% | 818 | 2136 | 2794 | 3167 | 3252 | 12167 | 10.9 | |
| | | Unfunded | Gold | 362 | 7.5% | 230 | 753 | 956 | 1087 | 1066 | 4092 | 11.3 | 106% |
| | | | Hybrid Gold | 78 | 1.6% | 112 | 122 | 213 | 172 | 188 | 807 | 10.3 | 88% |
| | | | Bronze | 189 | 3.9% | 102 | 268 | 378 | 415 | 442 | 1605 | 8.5 | 55% |
| | | | Green | 562 | 11.6% | 327 | 1122 | 1490 | 1621 | 1591 | 6151 | 10.9 | 99% |
| | | | Only Green | 163 | 3.4% | 97 | 290 | 391 | 441 | 418 | 1637 | 10.0 | 83% |
| | | | Paywalled | 1695 | 35.0% | 676 | 1712 | 2203 | 2352 | 2366 | 9309 | 5.5 | |
| Pediatrics | 1118 | Funded | Gold | 18 | 1.6% | 1 | 11 | 21 | 14 | 31 | 78 | 4.3 | -87% |
| | | | Hybrid Gold | 2 | 0.2% | 0 | 1 | 8 | 7 | 7 | 23 | 11.5 | 42% |
| | | | Bronze | 13 | 1.2% | 3 | 23 | 34 | 35 | 46 | 141 | 10.8 | 34% |
| | | | Green | 82 | 7.3% | 38 | 133 | 204 | 202 | 223 | 800 | 9.8 | 20% |
| | | | Only Green | 57 | 5.1% | 34 | 110 | 156 | 161 | 169 | 630 | 11.1 | 36% |
| | | | Paywalled | 74 | 6.6% | 38 | 83 | 129 | 173 | 177 | 600 | 8.1 | |
| | | Unfunded | Gold | 108 | 9.7% | 10 | 69 | 99 | 114 | 148 | 440 | 4.1 | -5% |
| | | | Hybrid Gold | 5 | 0.4% | 5 | 15 | 32 | 54 | 52 | 158 | 31.6 | 636% |
| | | | Bronze | 92 | 8.2% | 5 | 69 | 75 | 94 | 119 | 362 | 3.9 | -9% |
| | | | Green | 184 | 16.5% | 48 | 146 | 208 | 295 | 327 | 1024 | 5.6 | 30% |
| | | | Only Green | 67 | 6.0% | 33 | 60 | 87 | 141 | 149 | 470 | 7.0 | 63% |
| | | | Paywalled | 682 | 61.0% | 183 | 415 | 656 | 779 | 897 | 2930 | 4.3 | |
| Sensory Systems | 4708 | Funded | Gold | 724 | 15.4% | 329 | 1841 | 2428 | 2566 | 2604 | 9768 | 13.5 | 21% |
| | | | Hybrid Gold | 100 | 2.1% | 106 | 245 | 366 | 349 | 366 | 1432 | 14.3 | 28% |
| | | | Bronze | 246 | 5.2% | 212 | 456 | 535 | 618 | 583 | 2404 | 9.8 | -14% |
| | | | Green | 1190 | 25.3% | 842 | 3088 | 3899 | 4050 | 4233 | 16112 | 13.5 | 21% |
| | | | Only Green | 350 | 7.4% | 340 | 959 | 1113 | 1131 | 1251 | 4794 | 13.7 | 23% |
| | | | Paywalled | 522 | 11.1% | 444 | 1162 | 1378 | 1400 | 1449 | 5833 | 11.2 | |
| | | Unfunded | Gold | 648 | 13.8% | 255 | 1552 | 2144 | 2187 | 2418 | 8556 | 13.2 | 39% |
| | | | Hybrid Gold | 88 | 1.9% | 76 | 256 | 321 | 329 | 378 | 1360 | 15.5 | 62% |
| | | | Bronze | 348 | 7.4% | 183 | 556 | 761 | 812 | 938 | 3250 | 9.3 | -2% |
| | | | Green | 861 | 18.3% | 381 | 1747 | 2348 | 2475 | 2813 | 9764 | 11.3 | 19% |
| | | | Only Green | 178 | 3.8% | 79 | 291 | 357 | 348 | 380 | 1455 | 8.2 | -16% |
| | | | Paywalled | 1504 | 31.9% | 884 | 2753 | 3371 | 3621 | 3683 | 14312 | 9.5 | |
| Structural Biology | 5557 | Funded | Gold | 406 | 7.3% | 220 | 1140 | 1474 | 1646 | 1698 | 6178 | 15.2 | -5% |
| | | | Hybrid Gold | 202 | 3.6% | 268 | 861 | 1160 | 1387 | 1627 | 5303 | 26.3 | 64% |
| | | | Bronze | 473 | 8.5% | 447 | 1484 | 1785 | 1770 | 1867 | 7353 | 15.5 | -3% |
| | | | Green | 1377 | 24.8% | 1599 | 5097 | 6248 | 6452 | 7009 | 26405 | 19.2 | 20% |
| | | | Only Green | 571 | 10.3% | 885 | 2444 | 2878 | 2814 | 3026 | 12047 | 21.1 | 32% |
| | | | Paywalled | 1523 | 27.4% | 1298 | 4439 | 5996 | 6111 | 6491 | 24335 | 16.0 | |
| | | Unfunded | Gold | 464 | 8.3% | 257 | 949 | 1348 | 1460 | 1418 | 5432 | 11.7 | 23% |
| | | | Hybrid Gold | 53 | 1.0% | 34 | 142 | 163 | 157 | 149 | 645 | 12.2 | 28% |
| | | | Bronze | 154 | 2.8% | 105 | 326 | 371 | 387 | 410 | 1599 | 10.4 | 9% |
| | | | Green | 694 | 12.5% | 413 | 1414 | 1927 | 2016 | 1972 | 7742 | 11.2 | 17% |
| | | | Only Green | 144 | 2.6% | 93 | 271 | 341 | 321 | 309 | 1335 | 9.3 | -3% |
| | | | Paywalled | 1567 | 28.2% | 857 | 2869 | 3562 | 3721 | 3905 | 14914 | 9.5 | |
| Tourism, Leisure & Hospitality Management | 3491 | Funded | Gold | 12 | 0.3% | 0 | 8 | 17 | 20 | 30 | 75 | 6.3 | -198% |
| | | | Hybrid Gold | 22 | 0.6% | 17 | 74 | 129 | 137 | 173 | 530 | 24.1 | 29% |
| | | | Bronze | 19 | 0.5% | 15 | 63 | 98 | 120 | 128 | 424 | 22.3 | 20% |
| | | | Green | 114 | 3.3% | 72 | 266 | 443 | 613 | 719 | 2113 | 18.5 | -1% |
| | | | Only Green | 85 | 2.4% | 52 | 192 | 298 | 461 | 549 | 1552 | 18.3 | -2% |
| | | | Paywalled | 389 | 11.1% | 252 | 949 | 1523 | 2005 | 2525 | 7254 | 18.6 | |
| | | Unfunded | Gold | 116 | 3.3% | 11 | 93 | 147 | 166 | 231 | 648 | 5.6 | -160% |
| | | | Hybrid Gold | 33 | 0.9% | 32 | 97 | 162 | 261 | 303 | 855 | 25.9 | 78% |
| | | | Bronze | 9 | 0.3% | 0 | 8 | 17 | 26 | 27 | 78 | 8.7 | -67% |
| | | | Green | 476 | 13.6% | 245 | 894 | 1531 | 2231 | 2700 | 7601 | 16.0 | 10% |
| | | | Only Green | 393 | 11.3% | 227 | 778 | 1335 | 1954 | 2371 | 6665 | 17.0 | 17% |
| | | | Paywalled | 2413 | 69.1% | 1241 | 4113 | 7207 | 9729 | 12738 | 35028 | 14.5 | |
| Transplantation | 4076 | Funded | Gold | 26 | 0.6% | 7 | 26 | 53 | 60 | 50 | 196 | 7.5 | -36% |
| | | | Hybrid Gold | 217 | 5.3% | 285 | 893 | 989 | 1022 | 1128 | 4317 | 19.9 | 95% |
| | | | Bronze | 475 | 11.7% | 737 | 2202 | 2471 | 2561 | 2479 | 10450 | 22.0 | 115% |
| | | | Green | 660 | 16.2% | 978 | 2853 | 3310 | 3456 | 3384 | 13981 | 21.2 | 107% |
| | | | Only Green | 183 | 4.5% | 217 | 582 | 770 | 769 | 663 | 3001 | 16.4 | 60% |
| | | | Paywalled | 412 | 10.1% | 214 | 819 | 1079 | 1065 | 1037 | 4214 | 10.2 | |
| | | Unfunded | Gold | 202 | 5.0% | 39 | 193 | 239 | 241 | 228 | 940 | 4.7 | -54% |
| | | | Hybrid Gold | 143 | 3.5% | 184 | 521 | 550 | 613 | 633 | 2501 | 17.5 | 145% |
| | | | Bronze | 602 | 14.8% | 906 | 2533 | 3041 | 2772 | 2711 | 11963 | 19.9 | 178% |
| | | | Green | 369 | 9.1% | 443 | 1314 | 1490 | 1535 | 1542 | 6324 | 17.1 | 140% |
| | | | Only Green | 94 | 2.3% | 82 | 254 | 301 | 281 | 285 | 1203 | 12.8 | 79% |
| | | | Paywalled | 1722 | 42.2% | 636 | 2455 | 3324 | 3005 | 2888 | 12308 | 7.1 | |
| Visual & Performing Arts | 6729 | Funded | Gold | 18 | 0.3% | 1 | 20 | 23 | 24 | 28 | 96 | 5.3 | 51% |
| | | | Hybrid Gold | 17 | 0.3% | 4 | 15 | 18 | 26 | 21 | 84 | 4.9 | 40% |
| | | | Bronze | 11 | 0.2% | 5 | 9 | 16 | 24 | 24 | 78 | 7.1 | 101% |
| | | | Green | 101 | 1.5% | 40 | 79 | 111 | 144 | 159 | 533 | 5.3 | 49% |
| | | | Only Green | 68 | 1.0% | 33 | 51 | 63 | 86 | 97 | 330 | 4.9 | 37% |
| | | | Paywalled | 162 | 2.4% | 48 | 98 | 139 | 139 | 148 | 572 | 3.5 | |
| | | Unfunded | Gold | 333 | 4.9% | 19 | 37 | 74 | 63 | 93 | 286 | 0.9 | 38% |
| | | | Hybrid Gold | 93 | 1.4% | 7 | 38 | 59 | 89 | 90 | 283 | 3.0 | 388% |
| | | | Bronze | 282 | 4.2% | 9 | 48 | 77 | 68 | 103 | 305 | 1.1 | 74% |
| | | | Green | 817 | 12.1% | 97 | 302 | 423 | 475 | 540 | 1837 | 2.2 | 261% |
| | | | Only Green | 502 | 7.5% | 79 | 242 | 315 | 358 | 391 | 1385 | 2.8 | 343% |
| | | | Paywalled | 5243 | 77.9% | 218 | 628 | 728 | 815 | 878 | 3267 | 0.6 | |